\documentclass[a4paper,fleqn]{cas-dc}

\usepackage[numbers]{natbib}
\usepackage{graphicx}
\usepackage{amsmath,amssymb}
\usepackage{algorithm}
\usepackage{algpseudocode}
\usepackage{booktabs}
\usepackage{multirow}
\usepackage{makecell}
\usepackage{xcolor}
\usepackage{diagbox}
\usepackage{caption}
\usepackage{booktabs}
\usepackage{tabularx}
\usepackage{array}
\usepackage{makecell}
\usepackage{booktabs}
\usepackage{tabularx}
\usepackage{array}
\usepackage{amssymb}
\usepackage[section]{placeins}
\usepackage{placeins}
\usepackage{float}
\usepackage{stfloats}

\newcommand{\tblcheck}{\ensuremath{\checkmark}}

\newcolumntype{L}{>{\raggedright\arraybackslash}X}

\graphicspath{{img/}}

\def\tsc#1{\csdef{#1}{\textsc{\lowercase{#1}}\xspace}}
\tsc{WGM}
\tsc{QE}

\begin{document}
\let\WriteBookmarks\relax
\def\floatpagepagefraction{1}
\def\textpagefraction{.001}
\newcommand{\epnco}{EP-NCO}
\newenvironment{acknowledgements}{
  \section*{Acknowledgements}
}{}
\shorttitle{}    

\shortauthors{}  

\title[mode=title]{\epnco: Latency-Aware Service Placement using Neural Combinatorial Optimisers for Edge-Cloud Systems}



\author[1]{Kimia Abedpour}
[orcid=0009-0009-5961-4633,]
\cormark[1]
\ead{kabedpour01@qub.ac.uk}
\credit{
Conceptualization, Methodology, Software, Formal analysis, Investigation, Writing – original draft}

\author[2]{Mohammadsadeq Garshasbi Herabad}
[orcid=0000-0002-2336-2077,]
\ead{mohammadsadeq.garshasbi.herabad@kau.se}
\credit{
Formal analysis, Validation, Writing – review \& editing}
\author[1]{Zheng Li}
[orcid=0000-0002-9704-7651,]
\ead{zheng.li@qub.ac.uk}
\credit{Supervision, Writing – review \& editing}
\author[1,2]{Javid Taheri}
[orcid=0000-0001-9194-010X]
\ead{j.taheri@qub.ac.uk}

\credit{
Supervision, Methodology, Writing – review \& editing}

\affiliation[1]{
  organization={School of Electronics, Electrical Engineering and Computer Science},
  organization2={Queen’s University Belfast},
  addressline={},
  city={Belfast},
  postcode={BT7 1NN},
  country={UK}
}

\affiliation[2]{
  organization={Department of Mathematics and Computer Science},
  organization2={Karlstad University},
  city={Karlstad},
  postcode={651 88},
  country={Sweden}
}


\cortext[1]{Corresponding author}
\fntext[1]{}

\shorttitle{Latency-Aware Neural Service Placement}
\shortauthors{Abedpour et al.}

\begin{abstract}
Latency-sensitive Internet of Things (IoT) applications are increasingly deployed as interdependent microservices across heterogeneous edge--cloud infrastructures. Placing them under resource and network constraints is an NP-hard combinatorial optimisation problem. This study proposes \epnco, a graph-based neural combinatorial optimisation framework for service placement in compute continuum platforms. The objective accounts for execution latency, direct-link communication delay, and an approximate sharing of bandwidth among dependency flows that coexist in the resulting placement. Across held-out graph sizes generated by the same synthetic framework, the \epnco~variant trained on Large instances with hard decoding achieved the lowest mean aggregate latency cost at every scale. It was 6.0\%--12.5\% lower than a genetic algorithm (GA), 14.4\%--20.8\% lower than particle swarm optimisation (PSO), and 1.2\%--9.4\% lower than the strongest reinforcement learning ablation. On Small instances, its mean cost was 10.5\% below the best Gurobi solutions found and close to Gurobi's mean lower bound. The observed inference time of the \epnco~variants was 0.88–1.03 s per instance, compared with 85–97 s for GA and PSO. As a result, the one-off training cost of the \epnco~variant trained on Large instances with hard decoding is recovered after roughly 1,000 placements.
\end{abstract}


\maketitle


\section{Introduction}
\label{introduction}

IoT technologies have enabled latency-sensitive applications such as smart cities, industrial automation, and intelligent transportation systems. However, the resulting scale and heterogeneity create design and management challenges, making scalable and maintainable software architectures essential \cite{supraja2025ai}. Microservice architectures address these challenges by decomposing applications into loosely coupled services, accelerated by lightweight containerisation technologies that reduce resource overhead and simplify orchestration \cite{taleb2025survey}.

Cloud computing offers scalable computing and networking resources; however, it suffers from high latency because of the distance between data centres and edge-connected devices. Edge computing mitigates this by extending computation closer to the data sources \cite{kong2022edge}. Consequently, edge--cloud infrastructures are distributed and heterogeneous in computing capacity and network connectivity, making efficient, adaptive service placement challenging under dynamic workloads and resource constraints. 

Service placement in edge--cloud environments is a combinatorial optimisation problem \cite{jiang2020neural}, which determines the optimal mapping between services and heterogeneous computing nodes while satisfying the resource, latency, and cost constraints. As microservice components and candidate nodes increase, the solution space grows exponentially, making exhaustive search infeasible. This problem is classified as NP-hard \cite{10.1145/3391196,malazi2022dynamic,sonkoly2021survey}, implying that optimal solutions cannot be guaranteed in polynomial time. This mandates approaches \cite{electronics15010065} that rely on rule-based heuristics, metaheuristics, or learning-based methods to explore the solution space and obtain high-quality placement strategies within a reasonable time.

Rule-based heuristic approaches \cite{mahjoubi2021optimal,khan2022code,wu2021online} emphasise low computational overhead and fast decision-making; however, they often sacrifice solution quality. Metaheuristic methods \cite{herabad2025psoga, apat2024hybrid, bey2024quantum, huang2020ant} explore the search space extensively and achieve higher-quality solutions, but at a significant computational cost, limiting their applicability to offline optimisation. Learning-based approaches \cite{fahimullah2024machine,sharma2024intelligent, lingayya2024dynamic} shift this cost to offline training and enable fast inference; however, they often suffer from high training complexity and limited generalisation under dynamic conditions.

Reinforcement learning (RL), particularly Q-learning, has been applied to decision-making in edge--cloud environments for adaptability to workload and resource variations \cite{clifton2020q, jang2019q}. However, the growth in state--action spaces limits its scalability and training efficiency in large-scale service placement scenarios \cite{liu2022deep, chen2022joint}.

\begin{figure}[]
\centering    
\includegraphics[width=\linewidth]{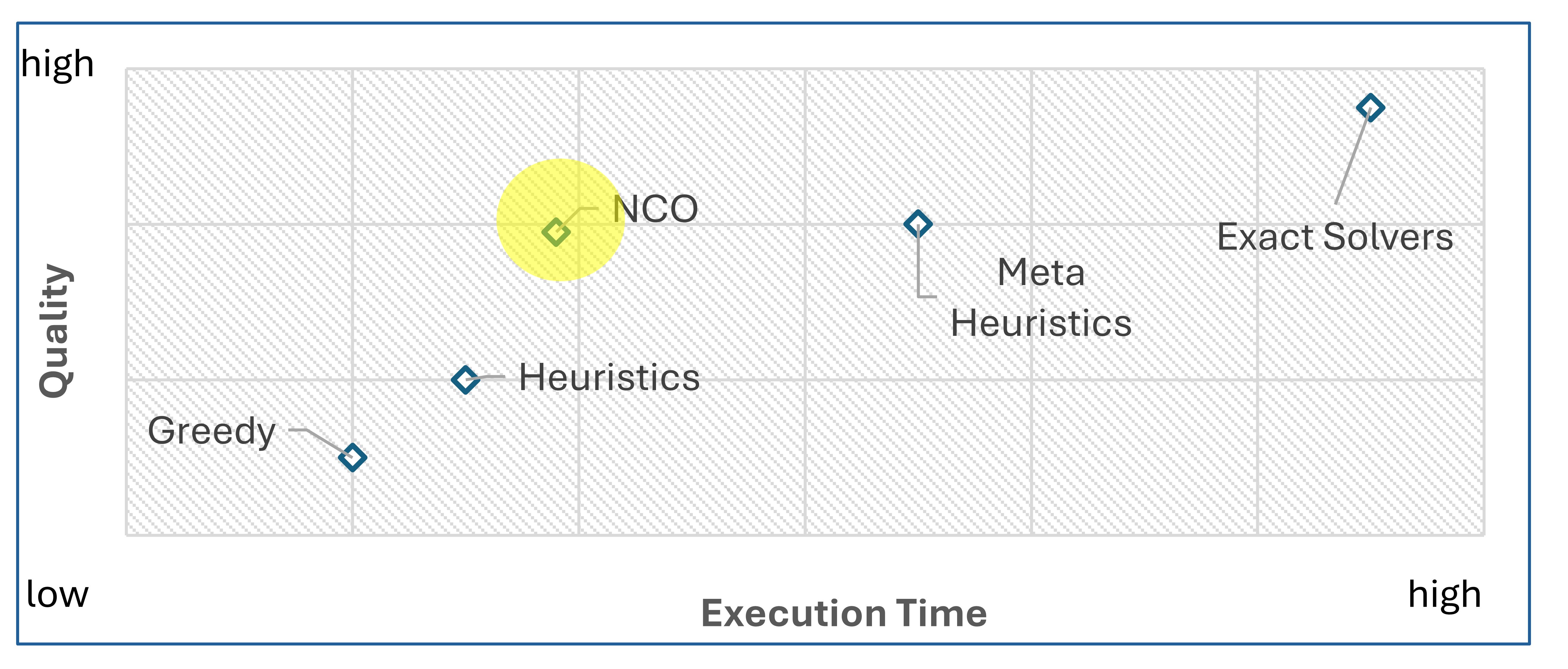}    
\caption{Conceptual comparison of solution quality versus online computation time for greedy, rule-based heuristic, metaheuristic, exact, and NCO approaches. The relative positions are qualitative and not drawn to scale.}
\label{fig:nco}
\end{figure}

Neural combinatorial optimisation (NCO) has emerged as a paradigm for solving large-scale combinatorial problems using neural networks. Typically, a neural policy encodes the problem instance and constructs a solution step by step, choosing one decision at a time (e.g., which node hosts the next service), and is trained to favour higher-quality solutions. By learning heuristics directly, often through RL, NCO can generate high-quality solutions without exhaustive search \cite{bengio2021methodological, vesselinova2020learning}. This makes it suitable for service placement problems with complex search spaces. Nevertheless, NCO-based solutions are in their infancy and face challenges, including generalisation across environments and the ability to handle strict constraints when solving multifaceted problems, such as service placement for distributed systems \cite{chung2025neural}. Figure~\ref{fig:nco} provides an illustration of the trade-off between solution quality and online computation time for optimisation paradigms. It is a qualitative schematic rather than an empirical comparison; relative positions are not to scale and do not establish a ranking.

We designed \epnco, which models the edge--cloud infrastructure as a graph of resource nodes and network links, and uses GNNs and RL to learn policies that map system states and service requirements to placement decisions that minimise aggregate latency under placement constraints. The contributions are as follows.

\begin{itemize}
    \item Developing a graph-based formulation of the service placement problem that captures infrastructure topology, heterogeneous node characteristics, and service requirements to provide a structured and scalable representation of edge--cloud systems.
    
    \item Proposing an NCO-based approach (\epnco) that integrates GNNs with reinforcement learning to learn efficient service placement policies directly from system states.
    
   \item Providing cross-size comparison of hard- and soft-decoding variants against rule-based heuristics, metaheuristics, and RL ablations on held-out instances, covering latency cost, feasibility, inference time, and amortisation of offline training cost.
\end{itemize}
The paper is organised as follows. Section~\ref{sec:related_work} reviews related work. Section~\ref{sec:System Model} describes the proposed system model and formulates optimisation objectives. Section~\ref{sec:methodology} details the design of the proposed solution (\epnco). Section~\ref{sec:experimental} presents the experimental setup and implementation details. Section~\ref{sec:results} evaluates the performance of all methods. Section~\ref{sec:conclusion} concludes the paper by summarising the findings and outlining future research directions. 

\section{Related Work}
\label{sec:related_work}

The placement of microservice applications across the compute continuum differs from conventional task offloading. Whereas an offloading decision typically determines where an individual task, or a task DAG, should execute, microservice placement must produce a reusable deployment whose interacting components compete for heterogeneous node and network resources. The resulting performance therefore depends not only on source--destination proximity, but also on the application call graph, the execution demands of each component, and the traffic induced between the selected hosts. Surveys of edge placement accordingly identify dependency awareness, runtime adaptation, and scalability as coupled rather than independent concerns~\cite{pallewatta2023taxonomy}. Existing approaches may be broadly classified into exact methods, heuristics, metaheuristics, and machine learning-based methods, the last of which encompass reinforcement learning (RL), graph neural networks (GNNs), and neural combinatorial optimisation (NCO).

\textbf{\textit{{Exact Methods}}} formulate placement as a mathematical or constraint-satisfaction problem and search the solution space systematically. Brogi and Forti modelled application components, infrastructure capabilities, and QoS constraints, and employed a backtracking search to identify eligible fog deployments~\cite{brogi2017qos}. 
Such formulations make component dependencies and resource constraints explicit and, where the search completes, guarantee feasibility or optimality with respect to the stated model. However, the size of the search space grows combinatorially with the number of components and nodes, and the reported evaluations are consequently confined to relatively small application--infrastructure instances.

\textbf{\textit{{Heuristic}}} approaches construct placements using problem-specific rules, such as greedy assignment to the node with the lowest expected latency, first-fit or best-fit resource matching, or the ranking of candidate nodes according to proximity and available capacity~\cite{gupta2017ifogsim, mahmud2018latency, taneja2017resource}. 
These methods produce decisions quickly and scale readily to large infrastructures, which renders them attractive for online operation. Their quality, however, depends on the design of the underlying rules, which are typically tailored to a particular objective or deployment scenario.

\textbf{\textit{{Metaheuristics}}} trade the guarantees of exact methods for tractable, near-optimal search. Pallewatta \emph{et al.} formulated batch microservice placement as a lexicographic combinatorial problem and applied set-based particle swarm optimisation (PSO) to balance makespan, cost, throughput, and resource utilisation~\cite{pallewatta2022qos}. In subsequent work, they proposed a reliability-aware formulation that combines PSO and NSGA-II with redundant placement in order to tolerate correlated fog failures~\cite{pallewatta2024reliability}. These studies demonstrate that component dependencies and infrastructure heterogeneity can be incorporated into rich multi-objective QoS models. Nevertheless, the population-based search must be repeated for every new problem instance, which makes such methods well suited to proactive or periodic planning.

\textbf{\textit{{Machine Learning-Based Methods}}} amortise the cost of search by training a policy offline and reusing it at inference time.

\textit{Reinforcement learning.} Liu \emph{et al.} jointly learned online service placement and computation-resource allocation using deep RL~\cite{liu2022deep}, whilst Afachao \emph{et al.} applied policy-gradient RL to microservice placement and resource management in an edge--cloud environment~\cite{afachao2024efficient}. In the compute-continuum setting, HephaestusForge learns multi-objective Kubernetes placement policies over latency, deployment cost, and distribution inequality, and its DeepSets representation permits transfer between cluster sets of differing cardinality~\cite{santos2025hephaestus}. Khansari and Sharifian formulated serverless microservice chaining as a partially observable Markov decision process in order to adapt placement under incomplete information~\cite{khansari2024scalable}.

\textit{Graph neural networks.} A further line of work incorporates relational structure into the learned policy. Lv \emph{et al.} embedded multiple microservice call graphs and employed graph RL to produce dependency-aware deployments~\cite{lv2023graph}. Chen \emph{et al.} combined a GNN with deep RL for deployment across cooperative edge nodes, allowing resource and invocation relations to inform the policy~\cite{chen2024gnnaided}. In closely related work, the DG-DDPG method of Chen \emph{et al.} uses two GNNs to encode the microservice application and the mobile edge computing (MEC) network respectively, and jointly learns placement and fractional resource allocation, with capacity and application-completeness constraints introduced through regularisation~\cite{chen2024adaptive}.

\textit{Neural combinatorial optimisation.} NCO learns a constructive heuristic, in which a decoder builds a complete solution step by step, conditioned on the partial solution generated so far~\cite{bengio2021methodological,mazyavkina2021rlco}. Jiang \emph{et al.} applied this paradigm to energy-efficient offloading in mobile edge computing~\cite{jiang2020neural}, and Xiao \emph{et al.} extended it with preference-conditioned attention, enabling a single trained model to generate latency--energy trade-offs without retraining for each preference~\cite{xiao2025nco}.

\textbf{\textit{{Research Gaps:}}}

The foregoing review reveals several open issues. First, exact methods and metaheuristics require an iterative search for every problem instance, which becomes computationally prohibitive when placements must be regenerated frequently or when the joint service--infrastructure graph is large. Second, heuristic rules are hand-crafted for particular scenarios and do not generally capture the interaction between service dependencies and the network links they traverse. Third, RL-based schedulers that rely on vector- or set-based state representations do not explicitly encode the service DAG or link-level dependency traffic. Fourth, existing graph-based RL methods produce continuous actor--critic actions in which resource and connectivity constraints are penalised rather than strictly enforced, so feasibility of the resulting placement is not guaranteed; moreover, their evaluations do not establish whether solution quality and decision speed are preserved as both the infrastructure and service graphs vary substantially in size. Finally, NCO has so far been applied to edge computing primarily in the form of task offloading, and has not been extended to multi-service DAG placement over a heterogeneous multi-tier substrate, in which bandwidth is shared among dependencies routed over the same link. To address these gaps, this paper proposes \epnco.
\section{Service Placement and System Model}
\label{sec:System Model}

\begin{figure}[]
    \centering
    \includegraphics[width=\linewidth]{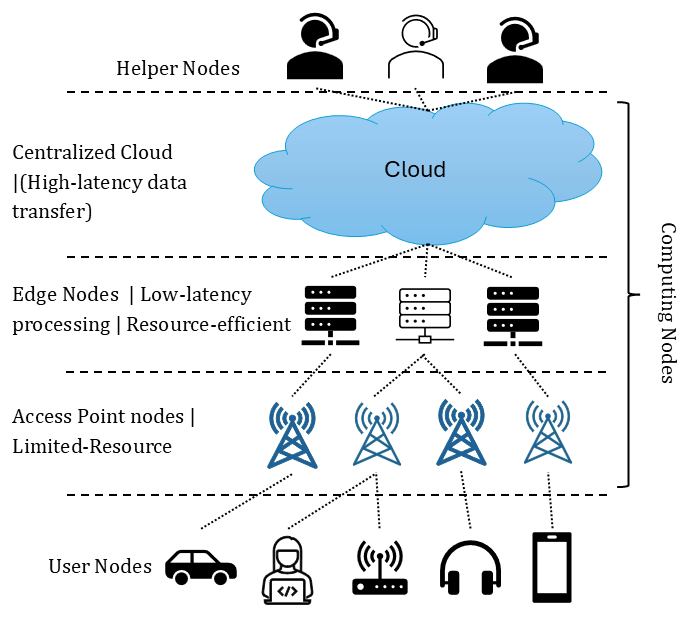}
    \caption{Multi-tier edge-cloud architecture including edge-connected devices, access point, edge, cloud, and cloud-connected devices layers.}
    \label{fig:system-arch}
\end{figure}

A multi-tier edge-cloud computing environment comprising heterogeneous computational and networking resources is considered to represent an edge-cloud system (the most common representation of compute continuum platforms) that spans from edge-connected devices to centralised cloud servers. As illustrated in Figure~\ref{fig:system-arch}, the architecture includes edge-connected devices, access points, edge nodes, cloud servers, and cloud-connected nodes. Collectively, they support the execution of distributed services to model a latency-aware service placement across the network.
\subsection{Service Placement Problem}
Service placement is a combinatorial optimisation problem in which service components are mapped to heterogeneous computing nodes under resource and performance constraints. Let $\mathcal{C}=\{c_1,\ldots,c_L\}$ denote the set of all service components (formally partitioned per service as $\mathcal{C}_s$, as detailed later in this section) and $\mathcal{N}=\{1,\ldots,N\}$ denote the set of available computing nodes. The binary decision variable $x_{c,n}\in\{0,1\}$ equals 1 if component $c$ is assigned to node $n$, and 0 otherwise.
Each instance is a static joint-placement snapshot containing multiple independent service DAGs that must be assigned within the same placement decision. CPU and memory requirements are simultaneous reservations: once a component is assigned, its reserved resources are not released within that instance. Equation~(\ref{eq:capacity}) enforces the resource capacity constraint on each node, ensuring that the total resource demand of assigned components does not exceed the available capacity:
\begin{equation}
\label{eq:capacity}
\sum_{c\in\mathcal{C}} x_{c,n}\mathbf{d}_c \le \mathbf{r}_n, \qquad \forall n\in\mathcal{N}
\end{equation}
Equation~(\ref{eq:obj}) defines the optimisation objective of the service placement problem, which seeks an assignment $\mathbf{x}$ that minimises the overall system cost, including \textit{execution link delay}, and \textit{latency cost}, formalised concretely later in this section as the total aggregate latency cost (Equation~\ref{eq:objective}):
\begin{equation}
\label{eq:obj}
\min_{\{x_{c,n}\}} f(\mathbf{x})
\end{equation}
Component dependencies within each service are modelled as a directed acyclic graph, as formalised later in this section (Equation~\ref{eq:service_dag}). Placing dependent service components on different nodes incurs communication costs and additional latency. The placement problem must satisfy strict CPU and memory capacity constraints and \textit{execution dependencies} while minimising performance costs, yielding an NP-hard optimisation problem with an exponentially large search space~\cite{taleb2025survey}.
\subsection{System Model}
\textbf{\textit{Computing Nodes Model:}}

The computing infrastructure is modelled as a directed graph representing a heterogeneous set of computational resources. Let $\mathcal{N}=\{1,\dots,N\}$ denote the set of computing nodes, which includes cloud nodes, edge nodes, access-point nodes, cloud-connected nodes, and edge-connected nodes.

Each node $n \in \mathcal{N}$ is characterised by its limited computational resources, namely its CPU capacity $\mathrm{cpuCap}_n$ (MIPS) and memory capacity $\mathrm{memCap}_n$ (MB).

\textbf{\textit{Network Model:}}

The network connectivity between nodes is represented as a directed graph $G^{net}=(\mathcal{N},\mathcal{E}^{net})$ where $(i,j)\in\mathcal{E}^{net}$ indicates the existence of a communication link from node $i$ to node $j$. Each link is associated with a bandwidth $\mathrm{bw}_{ij}>0$ and propagation link delay $\mathrm{delay}_{ij}$.


\textit{\textbf{Service Model:}}

Let $\mathcal{S}=\{1,\dots,S\}$ denote the set of services, where each service $s\in\mathcal{S}$ consists of a set of service components $\mathcal{C}_s$.

Equation~(\ref{eq:service_dag}) represents the internal dependency structure of each service as a directed acyclic graph:
\begin{equation}
\label{eq:service_dag}
G_s=(\mathcal{C}_s,\mathcal{E}_s)
\end{equation}
where an edge $(u,v)\in\mathcal{E}_s$ indicates that component $v$ depends on the output produced by component $u$ (i.e., $v$ can be executed after $u$ finishes).

Each component $c\in\mathcal{C}_s$ requires $\mathrm{cpuReq}_c$ (MI) units of CPU, $\mathrm{memReq}_c$ (MB) units of memory, and generates output data of size $\mathrm{dataSize}_c$ (MB).

\subsection{Objective Function}

The objective function Equation (\ref{eq:objective}) minimises the total latency cost of the deployed services, which is defined as the sum of the execution and communication link delays. This additive formulation models the aggregate latency cost.
\begin{equation}
\min_{\{x_{c,n}\}} \;
T_{\mathrm{total}}
=
T_{\mathrm{exec}}
+
T_{\mathrm{trans}}
\label{eq:objective}
\end{equation}

\textbf{\textit{Execution Time:}}

The execution time in Equation (\ref{eq:texec}) depends on the computational demand of each component and the processing capacity of the node on which it is deployed.  We interpret $\mathrm{cpuReq}_c$ in million instructions (MI) and $\mathrm{cpuCap}_n$ in million instructions per second (MIPS); therefore, their ratio is measured in seconds.
\begin{equation}
T_{\mathrm{exec}} =
\sum_{s\in\mathcal{S}}
\sum_{c\in\mathcal{C}_s}
\sum_{n\in\mathcal{N}}
x_{c,n}
\frac{\mathrm{cpuReq}_c}{\mathrm{cpuCap}_n}
\label{eq:texec}
\end{equation}

\textbf{\textit{Transmission Delay:}}

Communication link delays occur when dependent service components are deployed on different nodes. For each dependency $(u,v)\in\mathcal{E}_s$, the data generated by component $u$ must be transferred to the node hosting component $v$. 

To model bandwidth contention, let $s_{ij}$ denote the number of concurrent data flows traversing link $(i,j)$. Communication is restricted to direct network links, and multi-hop routing is not considered to maintain model tractability in large-scale settings. Accordingly, $s_{ij}$ denotes the number of service dependencies directly using link $(i,j)$.

Equation~(\ref{eq:effbw}) computes the effective bandwidth under bandwidth sharing among concurrent flows:
\begin{equation}
\mathrm{effBw}_{ij}=\frac{\mathrm{bw}_{ij}}{s_{ij}}
\label{eq:effbw}
\end{equation}
Equation (\ref{eq:ttrans}) calculates the total transmission link delay considering data transfer time and propagation link delay.

If no direct network link exists between the selected nodes, the placement is considered infeasible and excluded from the decision process.
\begin{equation}
T_{\mathrm{trans}} =
\sum_{s\in\mathcal{S}}
\sum_{(u,v)\in\mathcal{E}_s}
\left(
\frac{\mathrm{dataSize}_u}{\mathrm{effBw}_{a(u),a(v)}}
+
\mathrm{effLD}_{a(u),a(v)}
\right)
\label{eq:ttrans}
\end{equation}
\subsection{Constraints}

\textbf{\textit{Resource Capacity Constraints:}}

Constraints (\ref{eq:mem-cap}) and (\ref{eq:cpu-cap}) ensure that the total resource consumption of service components deployed on each node does not exceed the available node capacities.
\begin{equation}
\sum_{s\in\mathcal{S}}
\sum_{c\in\mathcal{C}_s}
x_{c,n}\,\mathrm{memReq}_c
\le
\mathrm{memCap}_n,
\quad \forall n\in\mathcal{N}
\label{eq:mem-cap}
\end{equation}
\begin{equation}
\sum_{c \in \mathcal{C}} x_{c,n}\,\mathrm{cpuReq}_{c}
\leq
\mathrm{cpuCap}_{n}\,\Delta,
\qquad \forall n \in \mathcal{N},
\label{eq:cpu-cap}
\end{equation}
where $\Delta$ denotes the scheduling interval in seconds and both sides of the constraint are measured in MI.

\textbf{\textit{Bandwidth Sharing Constraint:}}

When multiple dependency flows are mapped onto the same network link, the available bandwidth is shared among them according to the effective bandwidth definition in Equation (\ref{eq:effbw}).
\begin{table}[h]
\caption{System model-related notation.}
\label{tab:notation}
\centering

\setlength{\tabcolsep}{3pt}
\renewcommand{\arraystretch}{1.1}

{\fontsize{8pt}{9pt}\selectfont
\begin{tabularx}{\columnwidth}{
    >{\raggedright\arraybackslash}p{0.30\columnwidth}
    >{\raggedright\arraybackslash}X
}
\hline
\textbf{Notation} & \textbf{Description} \\
\hline

$\mathcal{N}$ &
Set of computing nodes (cloud, edge, access point,
cloud-connected devices, and edge-connected devices) \\

$\mathcal{E}^{\mathrm{net}}$ &
Set of directed network links between nodes \\

$G^{\mathrm{net}}
 = (\mathcal{N},\mathcal{E}^{\mathrm{net}})$ &
Network connectivity graph \\

$\mathrm{bw}_{ij}$ &
Bandwidth of link $(i,j)$ \\

$\mathrm{effLD}_{ij}$ &
Propagation link delay of link $(i,j)$ \\

$s_{ij}$ &
Number of concurrent flows sharing link $(i,j)$ \\

$\mathrm{effBw}_{ij}$ &
Effective bandwidth of link $(i,j)$ (shared) \\

\hline

$\mathcal{S}$ &
Set of services \\

$\mathcal{C}_{s}$ &
Set of service components in service $s$ \\

$G_s=(\mathcal{C}_s,\mathcal{E}_s)$ &
DAG representing the service structure \\

$(u,v)\in\mathcal{E}_s$ &
Dependency: component $v$ depends on the output of $u$ \\

$\mathrm{cpuReq}_{c}$ &
CPU requirement of component $c$ (MI) \\

$\mathrm{memReq}_{c}$ &
Memory requirement of component $c$ (MB) \\

$\mathrm{dataSize}_{c}$ &
Output data size of component $c$ (MB) \\

\hline

$\mathrm{cpuCap}_{n}$ &
CPU capacity of node $n$ (MIPS) \\

$\mathrm{memCap}_{n}$ &
Memory capacity of node $n$ (MB) \\

$x_{c,n}$ &
Binary placement variable (1 if component $c$ is placed
on node $n$) \\

$a(c)$ &
Node to which component $c$ is assigned \\

\hline

$T_{\mathrm{total}}$ &
Total latency cost (s) \\

$T_{\mathrm{exec}}$ &
Execution time of all service components \\

$T_{\mathrm{trans}}$ &
Transmission delay due to data transfers \\

\hline
\end{tabularx}
}

\end{table}

\section{\epnco: Proposed Solution}
\label{sec:methodology}
To solve the stated service placement problem, we designed the \epnco~framework to learn scalable placement policies that generate solutions for edge-cloud systems. \epnco~aims to address the inefficiency of exact solvers in relation to their exponentially growing execution time.

\textbf{\textit{Problem Statement (Service Placement):}}

The service placement problem is modelled as a combinatorial optimisation problem $\mathcal{P}=(X,f,\Omega)$, where $X\subseteq\{0,1\}^n$ represents the discrete solution space, $f:X\rightarrow\mathbb{R}$ denotes the objective function to be minimised or maximised, and $\Omega$ ensures that all constraints are satisfied, expressed as $g_i(x)\le0$ or $h_j(x)=0$. The optimal solution $x^{*}=\arg\min_{x\in X} f(x)$ is subject to these constraints.

\textbf{\textit{Neural Combinatorial Optimisation:}}

The \epnco~model learns solution-construction policies for discrete optimisation problems. Given a search space $X$ and objective $f:X\rightarrow\mathbb{R}$, it learns a parameterised policy $\pi_\theta:X\rightarrow\mathcal{A}$ that generates candidate solutions $x_\theta$ with low observed objective values. GNNs encode structural relationships in the problem instance, enabling the trained policy to generate a placement without an explicit online search procedure. This formulation does not guarantee that $x_\theta$ is optimal.

\textbf{\textit{Graph neural networks (GNNs):}}

GNNs operate on graph-structured data $\mathcal{G}=(\mathcal{V},\mathcal{E})$, where (a) each node $v_i$ has features $\mathbf{h}_i^{(0)}$ and (b) each edge encodes a communication or dependency relation. In this context, nodes correspond to either infrastructure or service graph nodes, depending on the graph type. GNNs are used to capture multi-hop structural dependencies to support downstream service placement tasks.

Equation~(\ref{eq:gnn_update}) defines the node representation update at layer $l$ in \epnco, where neighbour messages are aggregated:
\begin{equation}
\mathbf{h}_i^{(l+1)}=\phi^{(l)}\!\left(\mathbf{h}_i^{(l)},\;\bigoplus_{v_j\in\mathcal{N}(i)}\psi^{(l)}(\mathbf{h}_i^{(l)},\mathbf{h}_j^{(l)},\mathbf{e}_{ij})\right)
\label{eq:gnn_update}
\end{equation}

\subsection{Dual-Graph Representation and State Encoding}
\label{subsec:dual-graph}
\begin{figure*}[]
    \centering
    
    \includegraphics[width=\textwidth]{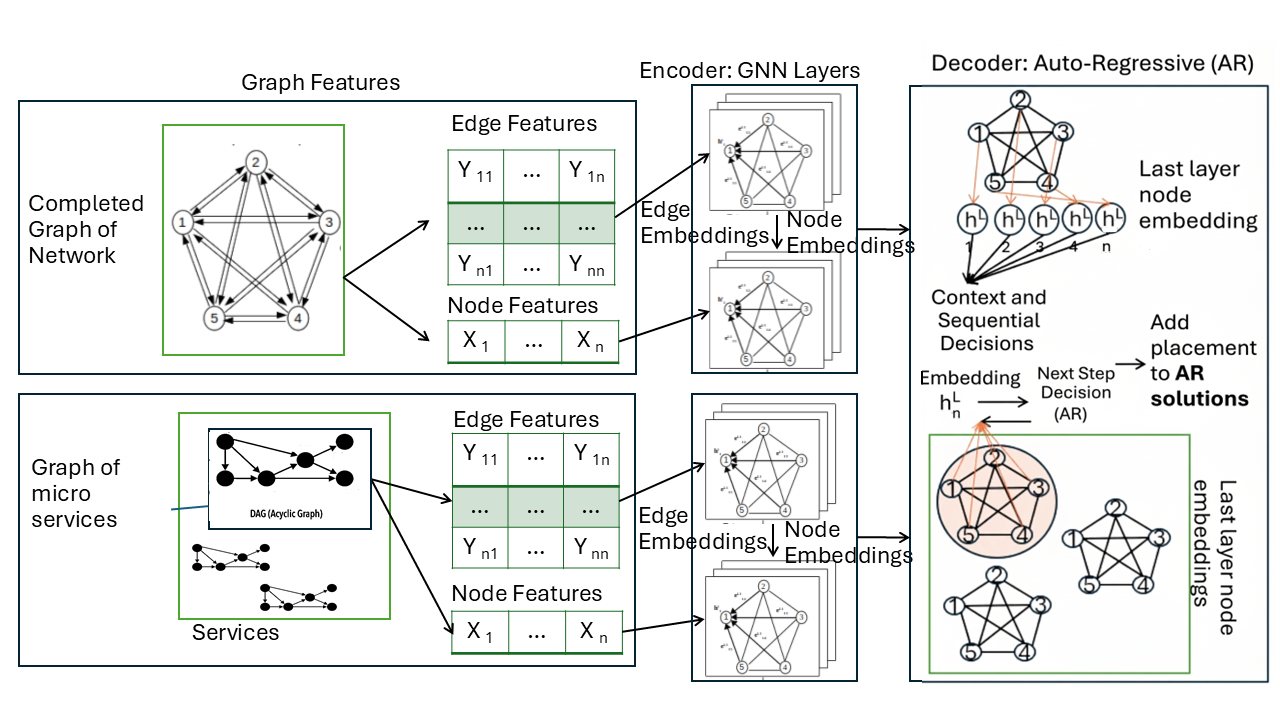}
    \caption{Dual-graph \epnco~architecture. Node graph uses bandwidth- and delay-aware messages, and Service graph uses data-size-gated messages. The implemented autoregressive decoder combines projected graph embeddings with placement features in a feed-forward scorer.}
    \label{fig:encoder-decoder}
\end{figure*}
Figure~\ref{fig:encoder-decoder} illustrates the NCO architecture of \epnco. The framework adopts a dual-graph representation modelling computing infrastructure and internal service structure. The encoder employs GNNs to extract structural embeddings from both graphs, whereas the decoder constructs a placement using an autoregressive policy. This design enables simultaneous reasoning about resource availability, network connectivity, and service-level dependencies.

\textbf{\textit{Network Graph Encoding:}}

The computing infrastructure is represented as a directed graph 
$G^{net}=(\mathcal{N},\mathcal{E}^{net})$, where the infrastructure graph nodes correspond to heterogeneous
computing entities throughout the compute continuum (cloud and edge-connected devices), and the infrastructure graph edges denote communication links. Each infrastructure graph node $n\in\mathcal{N}$ is associated with a feature vector describing its available computational resources (CPU and memory capacities), whereas each infrastructure graph edge $(i,j)\in\mathcal{E}^{net}$ captures link attributes between nodes (bandwidth and propagation link delay).

A GNN (\emph{NodeGNN}) encodes the infrastructure state. The infrastructure graph nodes are updated through a message-passing procedure in which each node aggregates information from its neighbouring infrastructure graph nodes. Equation~(\ref{eg:nco-update}) updates node $n$ at layer $k$ ($\mathbf{h}_n^{(k)}$) after each iteration.
\begin{equation}
\mathbf{h}_n^{(k+1)} =
\sigma
\left(
W^{(k)}
\cdot
\mathrm{AGG}
\left(
\{\mathbf{h}_u^{(k)} : u \in \mathcal{N}(n)\}
\right)
\right)
\label{eg:nco-update}
\end{equation}
Here, $\mathcal{N}(n)$ denotes the neighbourhood of node $n$, $W^{(k)}$ is a trainable weight matrix at layer $k$, $\mathrm{AGG}(\cdot)$ represents an aggregation function (e.g., mean aggregation), and $\sigma(\cdot)$ denotes a nonlinear activation function.

In \epnco, the NodeGNN comprises four message-passing layers and generates infrastructure graph node embeddings with a dimension of $32$ to capture both local resource availability and global network connectivity patterns. This provides a compact representation of the infrastructure state for the placement policy.

\textbf{\textit{Service Graph Encoding:}}

Each service is modelled as a DAG $G_s=(\mathcal{C}_s,\mathcal{E}_s)$, where the service graph nodes represent service components and service graph edges capture their data dependencies. Each service graph node is associated with a feature vector describing its computational requirements, including CPU demand, memory usage, and output data size.

Consistent with NodeGNN, another GNN (\emph{ServiceGNN}) encodes service DAGs to compute embeddings for all service graph nodes. Similar to NodeGNN, ServiceGNN performs iterative message passing along the edges of the service graph.

Equation~(\ref{eq:service_gnn}) defines the embedding update for a service graph node $c$ at layer $k$:
\begin{equation}
\mathbf{h}_c^{(k+1)} =
\sigma
\left(
W_s^{(k)}
\cdot
\mathrm{AGG}
\left(
\{\mathbf{h}_u^{(k)} : (u,c)\in\mathcal{E}_s\}
\right)
\right)
\label{eq:service_gnn}
\end{equation}
where $W_s^{(k)}$ denotes the trainable parameters of the ServiceGNN at layer $k$. This message-passing mechanism allows the encoder to capture the structural dependencies and precedence relations between service graph nodes.

The resulting embeddings summarise both the resource requirements and the structural role of each service graph node within the overall service workflow.

\subsection{Sequential Placement Model}
In this study, the service placement problem is solved using sequential service placement decisions, wherein service components are assigned to computing nodes one-by-one.

Equation~(\ref{eq:xcn}) defines the binary placement decision variable indicating whether component $c$ is assigned to node $n$:
\begin{equation}
x_{c,n} =
\begin{cases}
1, & \text{if component $c$ is assigned to node $n$}\\
0, & \text{otherwise}.
\end{cases}
\label{eq:xcn}
\end{equation}
Each component $c$ must be assigned to exactly one computing node $a(c)$ ($a(c)=n \quad \text{if } x_{c,n}=1$). During the placement process, service components are deployed sequentially in a topological order of each service DAG (Section~\ref{subsec:decoder}), so every predecessor is placed before its successors. At each decision step, a placement action selects a node for the next service component, considering resource availability and network feasibility constraints.

\subsection{Autoregressive Placement Policy and Constraint Handling}
\label{subsec:decoder}
Given the encoded dual-graph representation, \epnco~constructs placement solutions using an autoregressive decision process. Instead of optimising all service components to infrastructure nodes simultaneously, the model sequentially assigns one service component to one infrastructure node, thereby decomposing the original combinatorial problem into a series of conditional decision-making problems.

Let $\{c_1,c_2,\dots,c_C\}$ denote the deterministic sequence stored in each input file: service-list order followed by component-list order. For all generated instances, this stored order is a valid topological order of each service DAG (verified for every training and test instance), and it is not randomised between epochs. Consequently, when component $c_t$ is placed, all of its predecessors have already been assigned. At step $t$, the policy selects a host for $c_t$ from its observable decoder context $\mathbf{o}_t$ and previous assignments $\mathcal{A}_{<t}$.

\textbf{\textit{Soft Penalty-Based and Feasibility-Guided Hard Decoding Strategies:}}

To handle system constraints during the sequential placement process, \epnco~employs two decoding strategies: penalty-based \emph{soft-decoding} and \emph{feasibility-guided hard decoding}.

In the \textbf{soft-decoder} strategy, the autoregressive policy generates placement decisions without explicitly enforcing feasibility constraints during the decoding process. The policy assigns probabilities to all candidate infrastructure graph nodes, thereby allowing the model to explore a broader action space during training. Constraint violations, such as insufficient resource capacity or infeasible network connections, are not prevented during decision-making. Instead, they are incorporated into the optimisation objective through penalty terms added to the overall cost function. Consequently, infeasible solutions incur higher costs and are discouraged during learning.

In the \textbf{feasibility-guided hard-decoding strategy}, candidate node $n$ is retained at step $t$ when its residual CPU and memory are sufficient, applicable fixed-placement requirements are satisfied, 
and, for every predecessor of $c_t$, $n$ is either co-located with that predecessor or connected to it by an admissible direct logical edge. Because components are placed in topological order, this check covers every dependency edge of $c_t$. This is a step-wise action masking, not an exact global feasibility proof. When the admissible set is empty, the decoder first identifies CPU/memory-feasible nodes, relaxes the network and fixed-placement restrictions, and selects the node with the largest residual CPU. If no capacity-feasible node exists, it assigns the invalid marker $-1$, which is penalised by the objective. The branch is evaluated before probability construction, preventing an undefined softmax over a fully masked set.

The terms ``hard'' and ``soft'' refer to constraint handling, not deterministic versus stochastic action selection. During training, either decoder may sample actions, and during greedy evaluation, either may select the highest-scoring action.

Equation~(\ref{eq:autoregressive}) defines the factorisation of the joint assignment distribution under the placement policy:
\begin{equation}
\pi(\mathbf{x}) = \prod_{t=1}^{C} \pi\bigl(a(c_t)\mid \mathbf{o}_t,\mathcal{A}_{<t}\bigr)
\label{eq:autoregressive}
\end{equation}
The decoder context is
\begin{equation}
\mathbf{o}_t=\left(\mathbf{h}_{c_t},\mathbf{H}_n,\mathbf{r}^{cpu}_t,\mathbf{r}^{mem}_t,\mathbf{a}_{<t},\mathbf{q}^{net}_t,\mathbf{u}_t,\mathbf{m}_t\right),
\label{eq:decoder-context}
\end{equation}
where $\mathbf{h}_{c_t}$ is the current component embedding, $\mathbf{H}_n$ contains the node embeddings, $\mathbf{r}^{cpu}_t$ and $\mathbf{r}^{mem}_t$ are residual resources, $\mathbf{a}_{<t}$ contains previous assignments, $\mathbf{q}^{net}_t$ contains assignment-dependent predecessor bandwidth/delay and available dependency-data-size features, $\mathbf{u}_t$ contains the candidate crowding/utilisation features, and $\mathbf{m}_t$ is the mask. Link-flow counts and effective bandwidth are evaluated after the complete placement and are not dynamic policy features. We therefore describe an autoregressive constructive policy and do not rely on a formal minimal-state Markov proof.
\begin{table}[h]
\centering
\caption{Static and dynamic information boundaries in the
autoregressive policy.}
\label{tab:policy-information}

\setlength{\tabcolsep}{2pt}
\renewcommand{\arraystretch}{1.15}

{\fontsize{8pt}{9pt}\selectfont
\begin{tabularx}{\columnwidth}{LLccc}
\toprule

\textbf{Information} &
\textbf{Static/dynamic} &
\textbf{\makecell{Scor-\\ing}} &
\textbf{\makecell{Mask-\\ing}} &
\textbf{\makecell{Final\\objective\\only}} \\

\midrule

Component and node embeddings &
Static, precomputed &
Yes & No & No \\

Residual CPU and memory &
Dynamic &
Yes & Yes & No \\

Previous assignments &
Dynamic &
Yes & Yes & No \\

Predecessor bandwidth/delay and dependency size &
Assignment-dependent &
Yes & Yes & No \\

Crowding/utilisation features &
Dynamic candidate features &
Yes & No & No \\

Feasibility mask &
Dynamic &
No & Yes & No \\

Link-flow count and effective bandwidth &
Post-placement &
No & No & Yes \\

\bottomrule
\end{tabularx}
}

\end{table}
The decoder combines projected component and node embeddings with the dynamic candidate features in a feed-forward scoring function. Valid logits define a categorical distribution for stochastic training or a greedy argmax during evaluation.

\textbf{\textit{Step-Wise Feasibility-Guided Masking:}}

The hard decoder restricts inadmissible actions using the stepwise feasibility mask described above.

\textit{Resource Capacity Constraints}: A service component can only be assigned to an infrastructure node if sufficient residual CPU and memory resources are available. Infrastructure graph nodes that violate constraint~\eqref{eq:cpu-cap} or constraint~\eqref{eq:mem-cap} are masked and assigned a zero probability.

\textit{Connectivity Constraints}: For each dependency edge $(u,v)$ where service component $u$ has already been assigned to infrastructure node $i$, assigning service component $v$ to infrastructure node $j$ is allowed only if $j=i$ (co-location) or a corresponding communication link $(i,j)\in\mathcal{E}^{net}$ exists.

\textit{Incremental context updates}: After each valid assignment, residual CPU, residual memory, and the assignment vector are updated. Link-flow counts and effective bandwidth are calculated only after construction of the complete placement.

\textbf{\textit{Penalty-Based Handling of Residual Violations:}}

Certain global effects, such as bandwidth sharing among multiple simultaneous data transfers, can only be fully evaluated once a complete placement has been constructed. To capture these effects, \epnco~incorporates a penalty mechanism into its optimisation objective. Placements that induce severe link congestion, or that contain residual constraint violations, are penalised by adding a weighted violation term to the overall latency cost. This encourages the learned policy to generate placements that satisfy local feasibility constraints and achieve efficient global utilisation of network resources.

\textbf{\textit{Exploration and Inference Modes:}}

During training, the autoregressive policy samples node assignments according to the learned probability distribution, with temperature and entropy regularisation applied to promote exploration. For each problem instance, this produces one complete autoregressive rollout: the encoder embeddings are computed once, and the decoder then performs one conditional node-selection step for each service component. During evaluation, the policy uses a deterministic greedy rollout that selects the highest-probability node at each sequential step. This is not a single simultaneous prediction or a multi-start search for several candidate placements.

By combining stepwise feasibility-guided masking with penalty-based treatment of residual violations, the autoregressive decoder provides a scalable learning-based procedure for constructing placement solutions while explicitly accounting for the possibility of residual infeasibility.

\subsection{Reinforcement learning (RL) in \epnco}
In RL models, sequential decision-making is formulated as a Markov Decision Process (MDP) $\langle \mathcal{S},\mathcal{A},P,R,\gamma\rangle$, where a policy $\pi(a_t\mid s_t)$ selects actions to maximise the expected discounted return $G_t=\mathbb{E}\!\left[\sum_{k=0}^{\infty}\gamma^{k}r_{t+k}\right]$. By interacting with the environment, in this context, the agent learns placement policies that map states to actions without explicit searching. 

The service placement problem formulated in Section~\ref{sec:System Model} does not provide an explicit optimal supervision signal because the objective function depends on complex interactions between service component placements, resource contention, and network congestion effects. Moreover, exact optimisation is computationally infeasible for realistic system-scale applications. Consequently, \epnco~adopts the RL framework to learn an effective placement policy through direct interaction with the optimisation objective.

\textbf{\textit{Markov Decision Process Formulation:}}

The placement procedure is modelled as a finite-horizon MDP. At each decision step $t$, the agent observes the current system state, represented by the dual-graph embeddings described in Section~\ref{subsec:dual-graph}, and the residual resource and network states induced by previous assignments. The action at step $t$ selects a feasible infrastructure node for the current service component $c_t$, as detailed in Section~\ref{subsec:decoder}.

The environmental transition is deterministic and is defined by the updates following each assignment, including reductions in residual CPU and memory, updates to the assignment vector, and changes to the predecessor-based connectivity context. Link-flow counts and effective bandwidth are evaluated only after the complete placement has been constructed. An \textit{episode} terminates once all service components across all services are assigned. In \epnco, an \textit{episode} corresponds to constructing a complete placement solution for a given problem instance. Whereas an \textit{epoch} denotes a full pass over the training dataset comprising multiple episodes. In other words, each epoch consists of multiple episodes.

\textbf{\textit{Reward Design and Objective Alignment:}}

A reward signal is provided only after a complete placement solution is constructed. 
This episodic reward formulation directly reflects the global optimisation objective while avoiding biased intermediate signals that could distort the optimisation landscape.
The reward signal was derived directly from the optimisation objective. After the completion of a placement episode, the total latency cost is computed according to Equations \eqref{eq:texec} and \eqref{eq:ttrans}, incorporating both execution and transmission link delays, as well as bandwidth-sharing effects.

Equation~(\ref{eq:reward}) defines the episode-level reward as the negative total cost, thereby aligning the RL objective with the minimisation of the total latency cost:
\begin{equation}
R = -\bigl(T_{\mathrm{exec}} + T_{\mathrm{trans}} + \lambda \, V\bigr)
\label{eq:reward}
\end{equation}

where $V$ denotes the number of residual constraint violations, such as severe link congestion, and $\lambda$ is a penalty weight that controls the severity of these violations. This formulation ensures that the placement policy is explicitly discouraged from producing infeasible or highly congested placements while still allowing gradients to propagate during training.

\textbf{\textit{Policy Gradient Optimisation:}}

Equation~(\ref{eq:optobjective}) defines the optimisation objective for the placement policy with parameters $\theta$, including those of NodeGNN, ServiceGNN, and the autoregressive decoder. The objective of the learning process is to maximise the expected return obtained from the placement policies:
\begin{equation}
J(\theta)=
\mathbb{E}_{\pi_\theta}
\left[
R
\right]
\label{eq:optobjective}
\end{equation}

where $R$ denotes the episode-level reward, as defined in Equation~\eqref{eq:reward}.

The gradient of this objective is estimated using the \textit{REINFORCE} policy gradient estimator. Let $\{a_1,a_2,\dots,a_C\}$ denote the sequence of node selections generated by an autoregressive decoder.

Equation~(\ref{eq:rlgradient}) defines the policy gradient of the objective:
\begin{equation}
\nabla_\theta J(\theta)
=
\mathbb{E}_{\pi_\theta}
\left[
R
\sum_{t=1}^{C}
\nabla_\theta
\log
\pi_\theta(a_t|s_t)
\right]
\label{eq:rlgradient}
\end{equation}
where $s_t$ represents the system state at the decision step $t$. This gradient encourages action sequences that lead to lower latency costs and fewer violations of constraints.

In practice, a baseline $b$ is subtracted from the reward, as shown in Equation~(\ref{eq:rlgradient-baseline}):
\begin{equation}
\nabla_\theta J(\theta)
=
\mathbb{E}_{\pi_\theta}
\left[
\bigl(R-b\bigr)
\sum_{t=1}^{C}
\nabla_\theta
\log
\pi_\theta(a_t|s_t)
\right]
\label{eq:rlgradient-baseline}
\end{equation}
where $b$ is an instance-level baseline that does not depend on the sampled actions; it reduces the variance of the gradient estimate without biasing it.

Equation~(\ref{eq:entropy_obj}) defines the entropy-regularised optimisation objective, where entropy regularisation is incorporated to improve training stability:
\begin{equation}
J'(\theta)=
\mathbb{E}_{\pi_\theta}
\left[
R
\right]
+
\beta
\mathcal{H}(\pi_\theta)
\label{eq:entropy_obj}
\end{equation}
where $\mathcal{H}(\pi_\theta)$ denotes the policy entropy and $\beta$ controls the strength of the exploration regularisation.

\textbf{\textit{Exploration Strategy and Regularisation:}}

Effective exploration is critical because of the large and structured action spaces. During training, stochastic action selection samples the policy distribution at each decision step. A temperature parameter is applied to the action logits to control the distribution sharpness, thereby enabling a smooth transition from exploration to exploitation as training progresses.

Additionally, entropy regularisation is included in the optimisation objective to prevent premature convergence to suboptimal policies. The entropy coefficient is annealed during training, allowing for the initial exploration of diverse placement patterns before focusing on high-quality solutions.

\textbf{\textit{Inference and Generalisation:}}

At inference time, the trained policy constructs placements greedily, selecting the highest-probability node at each step, which yields deterministic and reproducible solutions.

\begin{algorithm}[h]
{\fontsize{8pt}{9pt}\selectfont
\caption{\epnco: Dual-Graph Encoder--Decoder with RL Optimisation}
\label{alg:NCO}
\begin{algorithmic}[1]
\Statex \textbf{Input:} Network graph $G^{net}=(\mathcal{N},\mathcal{E}^{net})$ with node capacities and link attributes
\Statex Service DAGs $\{G_s=(\mathcal{C}_s,\mathcal{E}_s)\}_{s\in\mathcal{S}}$
\Statex Penalty weight $\lambda$, epochs $E$, batch size $B$
\Statex Temperature schedule $\tau(\cdot)$, entropy schedule $\beta(\cdot)$
\Statex \textbf{Output:} Trained policy parameters $\theta$

 \State Initialise policy parameters $\theta$ (ServiceGNN, NodeGNN, AR-Decoder)

\For{epoch $=1$ to $E$}
    \State Sample a minibatch of $B$ placement instances
    \ForAll{instances in minibatch \textbf{(parallel)}}
        \State Initialise residual infrastructure node resources and link flow counters
        \State $\mathcal{A} \gets \emptyset$
        
        \State $\mathbf{H}_c \gets \mathrm{ServiceGNN}_{\theta}(\{G_s\})$
        \State $\mathbf{H}_n \gets \mathrm{NodeGNN}_{\theta}(G^{net})$
        
        \State Determine service component order $\{c_1,\dots,c_C\}$
        
        \For{$t=1$ to $C$}
            \State $c \gets c_t$
            
            \State Construct feasibility mask $\mathbf{m}\in\{0,1\}^{|\mathcal{N}|}$
            \State $\mathbf{m}[n]=1$ iff infrastructure node $n$ satisfies resource and connectivity constraints
            
            \State $\boldsymbol{\ell} \gets f_{\theta}(\mathbf{h}_c,\mathbf{H}_n)$
            \State $\boldsymbol{\ell}[n] \gets -\infty$ for all $n$ with $\mathbf{m}[n]=0$
            
            \If{training}
                \State $a(c)\sim \mathrm{softmax}(\boldsymbol{\ell}/\tau)$
            \Else
                \State $a(c)\gets \arg\max_n \boldsymbol{\ell}[n]$
            \EndIf
            
            \State $\mathcal{A}\gets \mathcal{A}\cup\{(c,a(c))\}$
            \State Update infrastructure node resources and link flow counters
        \EndFor
        
        \State Compute $T_{\mathrm{exec}}$, $T_{\mathrm{trans}}$ with bandwidth sharing
        \State $R \gets -(T_{\mathrm{exec}}+T_{\mathrm{trans}}+\lambda V)$
        \State Store log-probabilities and entropy
    \EndFor
    
    \State Update $\theta$ using policy gradient with entropy regularisation
\EndFor

\State \textbf{Inference:} run greedy decoding to obtain $\mathcal{A}$
\end{algorithmic}
}
\end{algorithm}

Overall, the proposed framework constructs candidate placements through stepwise masked or penalty-based decisions and evaluates complete-placement violations in the final objective~\cite{garg2021heuristicreinforcementlearningalgorithms,afachao2024efficient}.

\section{Experimental Setup}
\label{sec:experimental}

\textbf{\textit{Simulation Environment:}}

All experimental problem instances were generated using the edge-cloud simulation framework designed and introduced in \cite{electronics15010065,herabad2024optimizing}. This comprehensive simulator allows modelling of heterogeneous infrastructures with varying service-based workloads. To provide controlled heterogeneity, stochasticity, and reproducibility, all simulator parameters (including compute node capacities, network bandwidth, latency, component resource demands, and data transfer sizes) were sampled from predefined ranges derived from prior studies and industrial specifications. This enables realistic modelling of dynamic edge-cloud conditions and supports diverse deployment scenarios for both training and evaluation.
The implementation of the proposed \epnco~framework, including training and evaluation scripts, is publicly available at \cite{epnco_code}.

\textbf{\textit{Problem Instances:}}

Three instance scales (S:Small, M:Medium, L:Large) were created to assess performance under increasing structural complexity, with an additional XL:XLarge instance generated exclusively for scalability analysis. Each service comprised eight service components, presented as a DAG. The instance topologies included computing, edge-connected devices, and cloud-connected devices nodes. All simulations were executed in Python using PyTorch~2.9.0 (CPU) on Ubuntu~24.04.2~LTS with 14 CPU cores.
\begin{table}[h]
\caption{Problem Instance Specifications at Different System Scales}
\label{tab:instances}
\centering
\resizebox{\columnwidth}{!}{%
\setlength{\tabcolsep}{2.5pt}
\begin{tabular}{c c c c | c c}
\hline
\textbf{Scale} & \makecell{\textbf{Edge-}\\\textbf{connected}\\\textbf{devices}} & \textbf{Computing} & \makecell{\textbf{Cloud-}\\\textbf{connected}\\\textbf{devices}} & \textbf{Services} & \makecell{\textbf{Service}\\\textbf{components}} \\
\hline
S: Small   & 15 & 30  & 15 & 15 & 8 \\
M: Medium  & 30 & 60  & 30 & 30 & 8 \\
L: Large   & 50 & 100 & 50 & 50 & 8 \\
XL: XLarge & 75 & 145 & 75 & 75 & 8 \\
\hline
\end{tabular}
}
\end{table}

\subsection{Placement Algorithms}

\textit{\textbf{\epnco~setup and training details:}} We trained three \epnco~models on Small-, Medium-, and Large-scale synthetic instances, respectively. In a variant name, S, M, or L denotes this training scale, while H or S denotes hard or soft decoding. Each trained variant was evaluated on held-out Small, Medium, Large, and XLarge instances. Thus, ``cross-scale'' means transfer between graph sizes produced by the same simulator and parameter ranges; it does not mean transfer to a different workload distribution, simulator, or real deployment. Table~\ref{tab:trained-algos} summarises the naming convention.

The training instances included three categories of input features: (i) network-level attributes (bandwidth and link latency), (ii) node-level capacities (CPU and memory), and (iii) service-level requirements (CPU, memory, and output size). These features were encoded using NodeGNN and ServiceGNN encoders, and the model was trained using a policy gradient (REINFORCE) approach. For each instance, the policy constructs a complete placement through autoregressive decoding, after which the episode-level reward is computed and used for the policy-gradient update. \epnco~employs graph embeddings of dimension 32 and a four-layer GNN architecture for both the infrastructure and service encoders. 
The hyperparameters were kept fixed across all experiments.
 The data loader used \texttt{batch\_size=1}, whereas policy gradients were accumulated over 16 trajectories per update (four in CPU-only runs). The learning rates were $2\times10^{-4}$ for the encoders and $2\times10^{-3}$ for the decoder.

The reward is defined as the negative latency cost with a violation penalty weighted by $\lambda=1$ (Equation~\ref{eq:reward}). For variance reduction, the baseline $b$ is the negative GA latency cost of the instance when available, and an exponential moving average of past rewards otherwise. GA solutions are computed offline, so GA runtime is excluded from the reported \epnco~training time. The gradients were clipped to a maximum norm of 1.0. The temperature $\tau(\cdot)$ is linearly annealed from $0.8$ to $0.1$ over the first 60\% of training epochs and held fixed thereafter, while the entropy coefficient follows $\beta(p)=0.02e^{-3p}$ (with $p$ denoting normalised training progress), which decays towards a small positive floor of approximately $0.00099574$ rather than reaching zero, preserving a residual exploration incentive throughout the training. Early stopping monitors the mean latency cost on a validation set, evaluated using greedy decoding (selecting the highest-probability feasible action at each step), with a patience of eight evaluations and a minimum improvement threshold ($\mathrm{min\_delta}$) of $0.02$. All results were obtained using a fixed random seed and evaluated on four held-out test sets (Small, Medium, Large, and XLarge), each comprising exactly 200 placement instances loaded from a folder independent of the training data. Each training scale comprised 512 synthetically generated placement instances, which were sequentially loaded and processed (batchsize 1) by the data loader in each training epoch. These configurations were selected empirically to balance the training stability and computational efficiency of the available CPU-based hardware.

\textbf{\textit{Baseline Approaches:}} To evaluate \epnco, its performance was directly compared with several baseline approaches grouped into four categories according to their nature. First, the RL baselines were considered. Second, two widely used metaheuristic methods, namely, the genetic algorithm (GA) and particle swarm optimisation (PSO), were considered. Third, a set of rule-based heuristic strategies: Task Continuation Affinity (TCA), Most Powerful (MP), Most Data Size (MDS) and Least-Powerful (LP) \cite{electronics15010065,herabad2024optimizing}, and, finally, the exact method (Gurobi). The GA, PSO, and rule-based comparators follow the baseline implementations and server--client testbed methodology described by Herabad et al.~\cite{herabad2024optimizing}, which used Node.js, Docker, and Kubernetes; they were not reconstructed as new Python implementations for this study. The cited study reported that its workload data were validated in collaboration with an industry partner. In this study, GA and PSO were run on the same system and evaluated on the same problem instances, under the same objective function and constraints, as \epnco; the only difference is their implementation language (Node.js). A detailed comparative analysis of these approaches is presented in Section \ref{sec:results}.

\textbf{\textit{RL Baselines:}}
RL ablation variants, implemented within the proposed framework, were adapted from prior RL-based service placement studies \cite{garg2021heuristicreinforcementlearningalgorithms,afachao2024efficient} to match this paper's problem formulation (dual-graph state, autoregressive placement, feasibility masking, and latency-aware episodic reward), enabling a like-for-like comparison with \epnco. Published RL methods were not benchmarked directly, as their state-action spaces and objectives differ substantially. Both hard- and soft-decoder settings were evaluated: the hard decoder enforces feasibility during action selection, whereas the soft decoder penalises constraint violations.

\textbf{\textit{Genetic Algorithm}} is a population-based metaheuristic that iteratively evolves candidate placement solutions through selection, crossover, and mutation operators. It is particularly effective for multi-objective problems, such as minimising aggregate latency cost while maximising system reliability. Its iterative search incurs high computational overhead, and its runtime grows significantly with problem scale, limiting its suitability for real-time deployment \cite{electronics15010065,herabad2024optimizing}.

\textbf{\textit{Particle Swarm Optimisation}} is a swarm-intelligence metaheuristic in which candidate solutions are updated using both their individual best and the global best positions found so far. It can converge quickly towards high-quality solutions. However, its performance is sensitive to parameter tuning, and it scales poorly to large, complex search spaces \cite{electronics15010065,herabad2024optimizing}.

\textbf{\textit{Task Continuation Affinity}} is a rule-based heuristic that reduces communication overhead by placing dependent microservice components on the same or nearby nodes. It is computationally efficient and well-suited for online decision-making. However, despite these advantages, it relies on local optimisation and often fails to achieve globally optimal placements \cite{electronics15010065,herabad2024optimizing}.

\textbf{\textit{Most Powerful}} heuristic assigns microservice components to nodes with the highest available computational capacity to reduce execution time. However, it ignores network-related factors, such as communication link delay, which can lead to inefficient overall placements in distributed edge-cloud environments \cite{electronics15010065,herabad2024optimizing}.

\textbf{\textit{Least-Powerful}} heuristic selects nodes with the minimum sufficient resources required to host each component, aiming to improve resource utilisation. It is simple and effective in enhancing resource efficiency. However, despite these advantages, this approach can increase execution and communication link delays, often resulting in suboptimal performance for complex problem instances \cite{electronics15010065,herabad2024optimizing}.

\textbf{\textit{Most Data Size}} heuristic prioritises placing components that generate large output data on nodes with favourable network conditions to reduce transmission link delay. It is effective in minimising data transfer latency by considering communication patterns. However, despite its effectiveness in reducing transmission delay, its strong emphasis on data size often leads to poor performance in complex, large-scale placement scenarios \cite{electronics15010065,herabad2024optimizing}.

\label{subsec:trained-models}
\begin{table}[h]
\caption{Training configurations and algorithm names.}
\label{tab:trained-algos}
\centering

\setlength{\tabcolsep}{2pt}
\renewcommand{\arraystretch}{1.12}

{\fontsize{8pt}{9pt}\selectfont
\begin{tabularx}{\columnwidth}{
    >{\raggedright\arraybackslash}p{0.19\columnwidth}
    ccc
    cc
    >{\raggedright\arraybackslash}X
}
\toprule

\textbf{Name} &
\multicolumn{3}{c}{\textbf{Training scale}} &
\multicolumn{2}{c}{\textbf{Decoder}} &
\textbf{Type} \\

\cmidrule(lr){2-4}
\cmidrule(lr){5-6}

&
\textbf{Small} &
\textbf{Med.} &
\textbf{Large} &
\textbf{Hard} &
\textbf{Soft} &
\\

\midrule

\epnco\_SH &
\tblcheck & & & \tblcheck & &
 Proposed method \\

\epnco\_MH &
& \tblcheck & & \tblcheck & &
 Proposed method \\

\epnco\_LH &
& & \tblcheck & \tblcheck & &
 Proposed method \\

\epnco\_SS &
\tblcheck & & & & \tblcheck &
 Proposed method \\

\epnco\_MS &
& \tblcheck & & & \tblcheck &
 Proposed method \\

\epnco\_LS &
& & \tblcheck & & \tblcheck &
 Proposed method \\

\midrule

RL\_SH &
\tblcheck & & & \tblcheck & &
RL Ab \\

RL\_MH &
& \tblcheck & & \tblcheck & &
RL Ab \\

RL\_LH &
& & \tblcheck & \tblcheck & &
RL Ab \\

RL\_SS &
\tblcheck & & & & \tblcheck &
RL Ab \\

RL\_MS &
& \tblcheck & & & \tblcheck &
RL Ab \\

RL\_LS &
& & \tblcheck & & \tblcheck &
RL Ab \\

\midrule

GA~\cite{electronics15010065,herabad2024optimizing} &
\multicolumn{5}{l}{Genetic Algorithm baseline} \\

PSO~\cite{electronics15010065,herabad2024optimizing} &
\multicolumn{5}{l}{Particle Swarm Optimisation baseline} \\

TCA~\cite{electronics15010065,herabad2024optimizing} &
\multicolumn{5}{l}{Task Continuation Affinity heuristic} \\

MP~\cite{electronics15010065,herabad2024optimizing} &
\multicolumn{5}{l}{Most Powerful heuristic} \\

LP~\cite{electronics15010065,herabad2024optimizing} &
\multicolumn{5}{l}{Least-Powerful heuristic} \\

MDS~\cite{electronics15010065,herabad2024optimizing} &
\multicolumn{5}{l}{Most Data Size heuristic} \\
\bottomrule
\end{tabularx}

\vspace{2pt}

\parbox{\columnwidth}{%
\raggedright
 RL Ab= RL Ablation baseline, RL variants are adapted from
\cite{garg2021heuristicreinforcementlearningalgorithms,afachao2024efficient}.

}
}

\end{table}

\FloatBarrier
\section{Experimental Results \& Analyses}
\label{sec:results}

In this study, we evaluated the performance of \epnco~with the primary objective of reducing the aggregate latency cost while maintaining a low runtime, particularly in environments with limited resources. \epnco~performs initial training followed by inference, ensuring that it efficiently adapts to different operational conditions.

\subsection{Overall Performance Analysis}
\begin{figure}[]
    \centering
    \includegraphics[width=\linewidth]{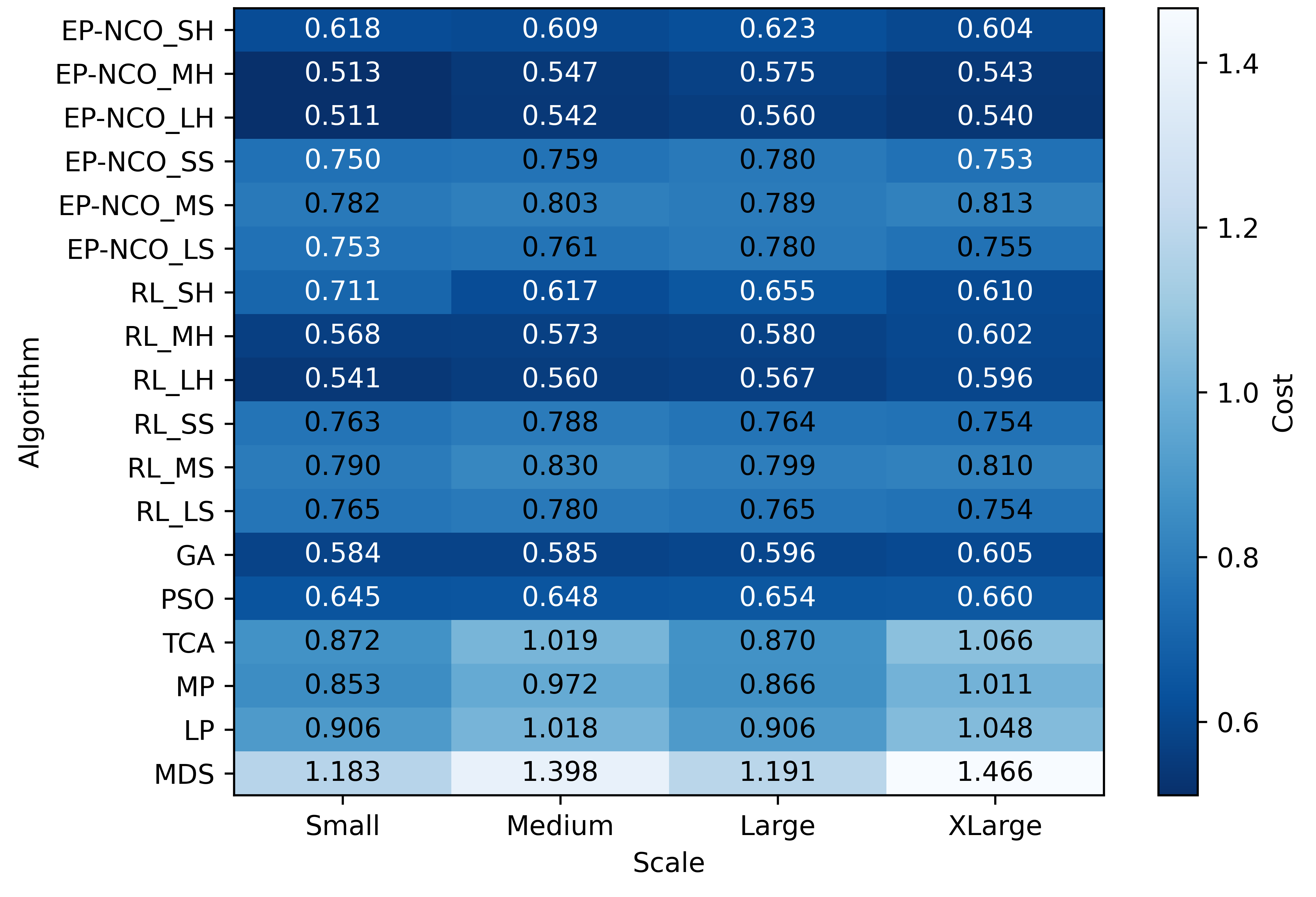}
    \caption{Mean aggregate latency cost (seconds per service) achieved by the evaluated algorithms across different infrastructure scales. Lower values indicate better performance.}
    \label{fig:heatmap}
\end{figure}
\begin{figure*}[b]
    \centering
    \includegraphics[width=\linewidth]{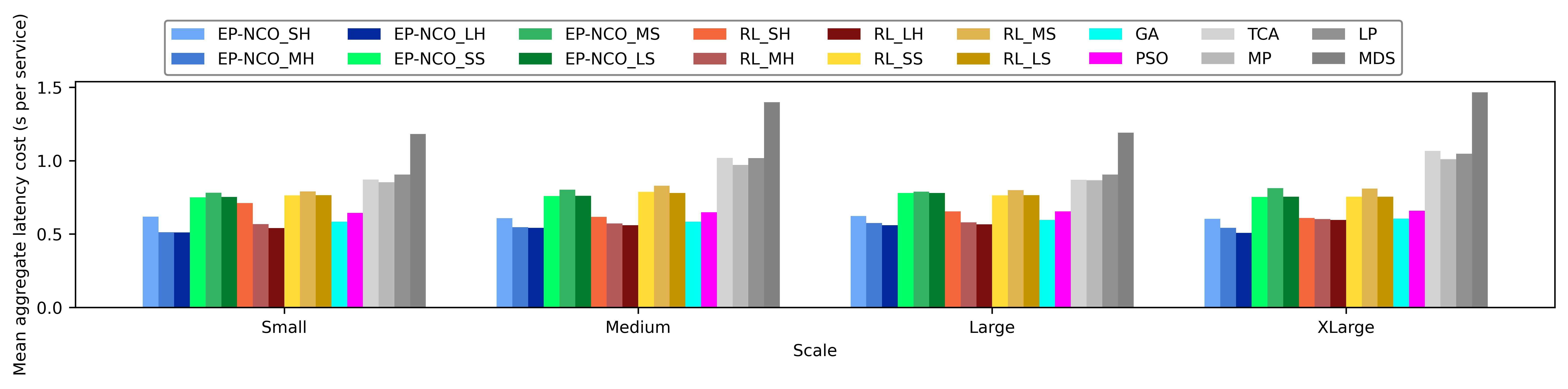}
    \caption{Grouped comparison of the mean aggregate latency cost (seconds per service) of the evaluated algorithms at each infrastructure scale. Lower values indicate better performance.}
    \label{fig:response}
\end{figure*}
Figure~\ref{fig:heatmap} compares the average aggregate latency cost achieved by the evaluated placement strategies across four infrastructure scales. Among all methods, \epnco\_LH consistently achieved the lowest latency cost, remaining between 0.511~s and 0.560~s per service across all scales (0.511~s at the Small scale and 0.540~s at the XLarge scale). \epnco\_MH also demonstrated competitive performance, although RL\_LH and RL\_MH achieved comparable results at several scales. In contrast, \epnco\_SH did not outperform all competing methods, as lower latency costs were obtained by \epnco\_LH, \epnco\_MH, RL\_LH, RL\_MH, and GA at every scale.

The soft-decoder \epnco~variants (SS, MS, and LS) produced higher latency costs than their hard-decoder counterparts and were generally outperformed by the GA and PSO. A similar distinction can be observed among the controlled RL ablation variants, where the hard-decoder configurations achieve lower latency costs than the corresponding soft-decoder models. GA and PSO remained relatively stable across the evaluated scales, with only modest increases in latency costs.

The rule-based heuristics (TCA, MP, LP, and MDS) incurred substantially higher latency costs, particularly at the Medium and XLarge scales. MDS consistently produced the highest latency cost, increasing from 1.183~s per service at the Small scale to 1.466~s per service at the XLarge scale. Overall, the results identify \epnco\_LH as the most effective placement strategy and demonstrate the performance advantage of hard decoding over soft decoding within the \epnco~family.

\subsection{Aggregate Latency Cost Analysis}

Figure~\ref{fig:response} presents aggregate latency cost results in seconds per service using a grouped bar chart. This representation makes the absolute performance differences among the algorithms more explicit. \epnco\_LH consistently achieved the lowest average latency cost across all infrastructure scales, generally followed by \epnco\_MH and the hard-decoder RL variants (at the Large scale, RL\_LH ranked second). GA and PSO demonstrated relatively stable performance but remained less effective than the strongest hard-decoder configurations.

The soft-decoder \epnco~and RL variants produced noticeably higher latency costs than the hard-decoder counterparts. The rule-based heuristics formed a clearly higher latency cost group, whereas MDS yielded the poorest performance at every scale. The differences became particularly pronounced at the XLarge scale, where MDS reached approximately 1.466~s per service, compared with 0.540~s per service for \epnco\_LH. These results reinforce the advantage of \epnco\_LH and show that the hard-decoder configurations trained on Medium and Large instances provide the most effective latency cost optimisation.
\subsection{Comparison with Gurobi Bounds}
\label{subsec:gurobi-comparison}

\begin{figure}[]
    \centering
    \includegraphics[width=\linewidth]{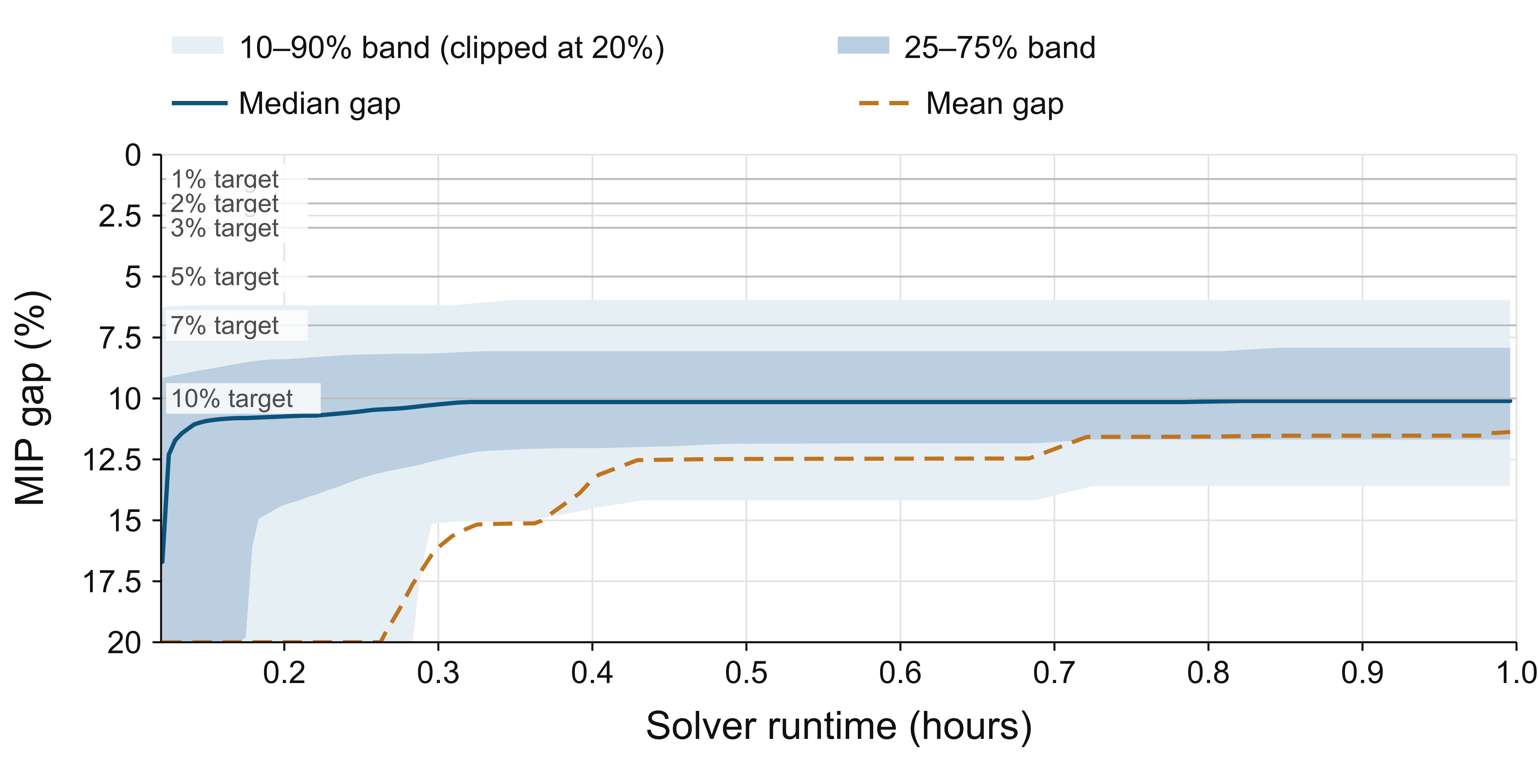}
    \caption{Evolution of the Gurobi MIP gap across the 200 Small-scale
    test instances. The lines show the smoothed median and mean gaps,
    while the shaded regions show the 25th--75th and 10th--90th
    percentile ranges. The outer percentile range is clipped at $20\%$
    to preserve visibility in the main region of the plot.}
    \label{fig:gurobi-gap-trend}
\end{figure}

To provide an exact optimisation reference for the proposed approach,
we evaluated Gurobi on all 200 Small-scale test instances, i.e., the
same instances used to evaluate \epnco~and the other baselines. Each
instance was solved independently using a standard Gurobi
branch-and-bound search with a time limit of one hour. Because the
considered problem is a minimisation problem, Gurobi maintains a
feasible incumbent objective $z^{\mathrm{inc}}$ and a valid lower bound
$z^{\mathrm{LB}}$. The reported Gurobi MIP gap is defined as
\begin{equation}
\mathrm{MIPGap}(\%) =
\frac{z^{\mathrm{inc}}-z^{\mathrm{LB}}}
{|z^{\mathrm{inc}}|}\times 100.
\label{eq:gurobi-mip-gap}
\end{equation}
\begin{table}[H]
\centering
\caption{Summary of the Gurobi results on the Small-scale test
instances.}
\label{tab:gurobi-summary}

\setlength{\tabcolsep}{2pt}
\renewcommand{\arraystretch}{1.1}

{\fontsize{8pt}{9pt}\selectfont
\begin{tabularx}{\columnwidth}{
*{7}{>{\centering\arraybackslash}X}
}
\toprule

\textbf{\makecell{Proven\\optimal}} &
\textbf{\makecell{Mean\\gap}} &
\textbf{\makecell{Median\\gap}} &
\textbf{\makecell{Gap $\leq$\\5\%}} &
\textbf{\makecell{Gap $\leq$\\10\%}} &
\textbf{\makecell{Gap $\leq$\\20\%}} &
\textbf{\makecell{Time\\limit}} \\

\midrule

0 & 10.61\% & 10.10\% & 14/200 & 98/200 & 196/200 & 3600\,s \\

\bottomrule
\end{tabularx}
}

\end{table}
The one-hour time limit was chosen empirically. As shown in
Figure~\ref{fig:gurobi-gap-trend}, the MIP gap had largely plateaued
well before the end of the one-hour period, and preliminary runs with
longer limits confirmed that the incumbent objective and MIP gap
improved only marginally beyond one hour. Additional runtime therefore
yielded little benefit. Under this limit, Gurobi found a feasible
placement for all instances. However, none of the instances was
solved to proven optimality, and all runs terminated with the
\texttt{TIME\_LIMIT} status. Therefore, the Gurobi solutions are
referred to as the \emph{best Gurobi incumbents}, rather than as
optimal solutions. As summarised in Table~\ref{tab:gurobi-summary},
the final mean and median MIP gaps were $10.61\%$ and $10.10\%$,
respectively. The final gap was at most $5\%$ for 14 instances, at
most $10\%$ for 98 instances, and at most $20\%$ for 196 of the 200
instances. The minimum and maximum final gaps were $3.12\%$ and
$58.46\%$, respectively.

Figure~\ref{fig:gurobi-gap-trend} shows the aggregate evolution of the
Gurobi MIP gap over the one-hour optimisation period. The solid line
shows the median gap, while the dashed line shows the mean gap. The
shaded regions represent the interquartile range and the 10th--90th
percentile range across instances. The median gap decreased rapidly
during the initial part of the optimisation, but its rate of
improvement became considerably smaller after approximately $0.3$
hours. This behaviour indicates that Gurobi could find feasible
placements relatively early, while subsequently tightening the
optimality bound was considerably more computationally demanding.

The bandwidth-sharing term makes the objective nonlinear, because the flow count $s_{ij}$ depends on the placement variables. To obtain a mixed-integer linear model for Gurobi, each product of placement variables was replaced by an auxiliary binary variable $y_{uv,ij}=x_{u,i}\,x_{v,j}$, linearised with the standard constraints $y_{uv,ij}\le x_{u,i}$, $y_{uv,ij}\le x_{v,j}$, and $y_{uv,ij}\ge x_{u,i}+x_{v,j}-1$.
Using the same per-service latency cost representation employed in
the remaining experiments, the Gurobi incumbents achieved an average
latency cost of $0.571$~s per service over the 200 instances, while
the corresponding Gurobi lower bounds had an average value of
$0.509$~s. Thus, the unknown average optimal objective lies between the mean Gurobi lower bound and the mean incumbent value. On the same 200
instances, \epnco\_LH achieved an average latency cost of $0.511$~s
per service. At the aggregate level, this represents a reduction of
approximately $10.5\%$ relative to the best feasible solutions found
by Gurobi within the one-hour time limit:
\begin{equation}
\Delta_{\mathrm{EP\mbox{-}NCO,G}}
=
\frac{0.511-0.571}{0.571}\times100
\approx -10.5\%.
\label{eq:nco-gurobi-difference}
\end{equation}

The \epnco\_LH average is also close to the average Gurobi lower bound
of $0.509$~s. This result provides evidence that the proposed method
can rapidly produce high-quality solutions in a region close to the
solver-derived lower bound, whereas Gurobi requires one hour per
instance and still retains a non-zero optimality gap. Nevertheless,
because Gurobi did not prove optimality for any of these instances,
its incumbents must not be interpreted as exact optimal solutions.
Furthermore, proximity between aggregate averages should be
interpreted as supporting evidence rather than as a certified mean
per-instance optimality gap.

Overall, these results highlight the complementary roles of the two
approaches. Gurobi provides valid lower bounds and an explicit measure
of residual optimality uncertainty, whereas \epnco~produces
high-quality placements in approximately one second per instance. The
comparison therefore demonstrates that \epnco~offers a favourable
trade-off between solution quality and computational efficiency for
latency-aware service placement.

\subsection{Algorithm Computational Time and Complexity Analysis}


Table~\ref{tab:ml_cost_profile} summarises the computational characteristics of each algorithm by reporting their training cost, inference latency, and cumulative CPU usage on the XL topology. The learning-based families (\epnco~and RL) incur substantial one-off training overheads, with variants trained on larger instances requiring significantly more computation, because each training episode involves more infrastructure nodes and service components. 

Despite these differences in the training cost, most learning models achieved sub-second inference times, indicating that once trained, they provide highly efficient per-request execution. Metaheuristics (GA and PSO) avoid training costs but exhibit considerably higher inference times. Rule-based heuristics (TCA, MP, LP, and MDS) remain computationally minimal but have proven ineffective for complex large-scale problem instances. 

Overall, the table highlights a fundamental trade-off: learning-based approaches (i.e., \epnco~and RL) pay a high upfront cost in exchange for extremely low inference latency, whereas non-learning-based methods (GA, PSO, TCA, MP, LP, and MDS) avoid training but incur higher per-execution runtimes with substantially lower solution quality for large-scale problem instances.
\begin{table}[H]
\centering
\caption{Algorithm runtimes and CPU cost on XLarge-scale instances.}
\label{tab:ml_cost_profile}

\setlength{\tabcolsep}{2pt}
\renewcommand{\arraystretch}{1.05}

{\fontsize{8pt}{9pt}\selectfont
\begin{tabularx}{\columnwidth}{
l
>{\centering\arraybackslash}X
c
>{\centering\arraybackslash}X
>{\centering\arraybackslash}X
}
\toprule

\textbf{Method} &
\textbf{\makecell{Train time\\(sec / h)}} &
\textbf{Cores} &
\textbf{\makecell{Inference\\(s)}} &
\textbf{\makecell{CPU cost\\(sec / h)}} \\

\midrule

\epnco\_SH & \makecell{9072\\(2.52 h)} & 14 & 0.92 & \makecell{127009\\(35.28 h)} \\
\epnco\_MH & \makecell{30672\\(8.52 h)} & 14 & 1.03 & \makecell{429409\\(119.28 h)} \\
\epnco\_LH & \makecell{95004\\(26.39 h)} & 14 & 0.97 & \makecell{1330056\\(369.46 h)} \\
\epnco\_SS & \makecell{10512\\(2.92 h)} & 14 & 0.88 & \makecell{147168\\(40.88 h)} \\
\epnco\_MS & \makecell{37800\\(10.50 h)} & 14 & 0.90 & \makecell{529200\\(147.00 h)} \\
\epnco\_LS & \makecell{87480\\(24.30 h)} & 14 & 0.90 & \makecell{1224720\\(340.20 h)} \\

RL\_SH & \makecell{11052\\(3.07 h)} & 14 & 0.80 & \makecell{154728\\(42.98 h)} \\
RL\_MH & \makecell{60120\\(16.70 h)} & 14 & 0.82 & \makecell{841680\\(233.80 h)} \\
RL\_LH & \makecell{117324\\(32.59 h)} & 14 & 0.89 & \makecell{1642536\\(456.26 h)} \\
RL\_SS & \makecell{4212\\(1.17 h)} & 14 & 0.88 & \makecell{58968\\(16.38 h)} \\
RL\_MS & \makecell{24120\\(6.70 h)} & 14 & 0.89 & \makecell{337680\\(93.80 h)} \\
RL\_LS & \makecell{50760\\(14.10 h)} & 14 & 0.78 & \makecell{710640\\(197.40 h)} \\

\midrule

GA  & 0 & 1 & 96.574 & 96.57 \\
PSO & 0 & 1 & 85.217 & 85.21 \\
TCA & 0 & 1 & 0.016 & 0.016 \\
MP  & 0 & 1 & 0.017 & 0.017 \\
LP  & 0 & 1 & 0.033 & 0.033 \\
MDS & 0 & 1 & 0.015 & 0.015 \\

\bottomrule
\end{tabularx}
}

\end{table}

\begin{figure*}[h]
    \centering
    \includegraphics[width=\linewidth]{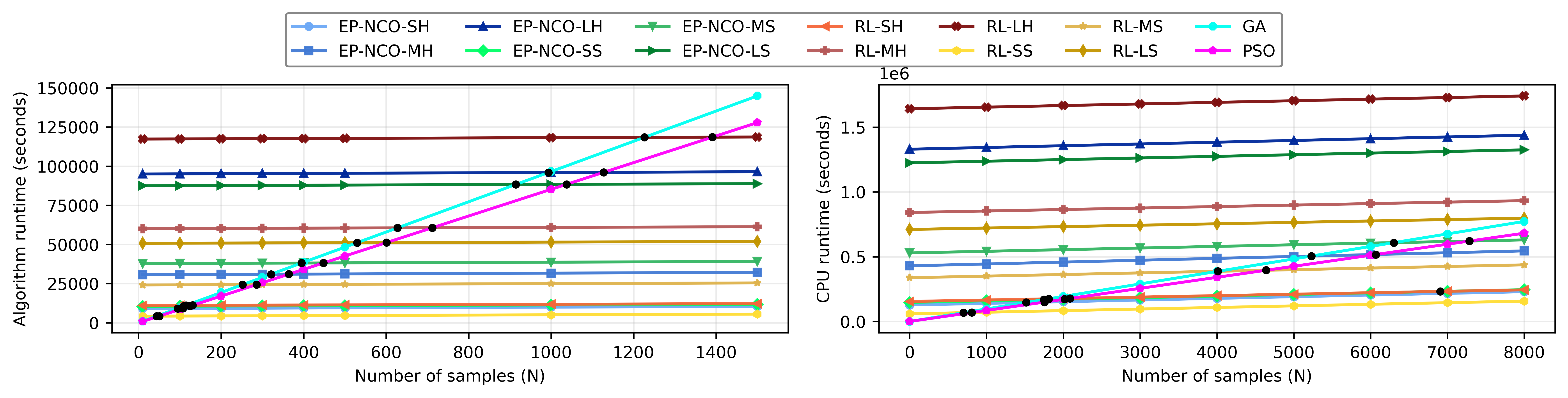}
    \caption{Crossover number of XL runs after which a learning method becomes faster end-to-end (training + N $\times$ inference), and crossover points by compute budget (CPU-hours). The learning methods are adjusted to accommodate the difference between 14 cores used in \epnco~vs. one core used in GA/PSO.}
    \label{fig:quality-time}
\end{figure*}

\textbf{\textit{Execution Time Analysis:}} 

In Figure \ref{fig:quality-time}, the runtime-based crossover curves illustrate the point at which the amortised cost of training a learning-based method becomes favourable relative to the repeated execution of non-learning-based approaches. For each learning algorithm, $N^*_{exetime}$ represents the number of problem instances that the algorithm must solve to compensate for its lengthy training time (as compared with another algorithm). Equation (\ref{eq:alg-exe-time}) is calculated by comparing the TotalAlgorithmExecutionTime (TAET)  for solving $N^*_{exetime}$ problem instances using algorithm $alg$. For example, both \epnco{\_SH} and GA take almost the same amount of time to solve $N^*_{exetime}=95$ problem instances because TAET(GA, $N^*_{exetime}$) = TAET(\epnco{\_SH}, $N^*_{exetime})$.
\begin{equation}
\label{eq:alg-exe-time}
 \text{TAET }(alg,N) = T_\mathrm{train}^\mathrm{alg} + N \times T_\mathrm{infer}^\mathrm{alg}   
\end{equation}

where $T_\mathrm{train}$ denotes the one-time training time, and $T_\mathrm{infer}$ represents the per-execution inference time. For the metaheuristics (GA and PSO), $T_\mathrm{train}=0$.

Because the inference costs of the \epnco~and RL models remain nearly constant across additional problem instances, their cumulative runtime grows slowly, whereas GA and PSO exhibit strictly linear growth driven by their expensive per-run optimisation cycles.

The crossover numbers are related to the algorithm nature. For example, \epnco\_SH surpasses GA after approximately 95 problem instances and PSO after 108 problem instances, whereas \epnco\_MH surpasses them after 321 and 364 problem instances, respectively. For the RL models, RL\_SH crossed GA at approximately 115 problem instances and PSO at 131 problem instances. This rapid amortisation of training costs highlights the efficiency of compact learning-based strategies for repeated large-scale placement decisions. Variants trained on larger instances (e.g. \epnco\_LH and RL\_LH) require higher crossover points owing to longer training times but ultimately achieve the same asymptotic advantage as the number of placements increases.

Overall, this analysis confirms that learning-based strategies become increasingly favourable in terms of the total wall-clock execution runtime when deployment decisions are repeatedly required.

\textbf{\textit{Core-Hour Analysis:}}

Core-hour analysis complements the runtime perspective by explicitly accounting for core utilisation and revealing how the total compute expenditure evolves using different algorithms. For each learning-based method, the crossover number of problem instances ($N^{*}_{cpu}$) at which the total cumulative CPU consumption surpasses that of another algorithm. Equation (\ref{eq:alg-cpu-time}) shows how the TotalAlgorithmCpuTime (TACT) is computed, and used to find the cross-over point ($N^{*}_{cpu}$) where the total core-hour of one algorithm surpasses another one.
\begin{equation}
\text{TACT}(\mathrm{alg}) = CT\!\left(T_\mathrm{train}^\mathrm{alg}\right) + N^{*}_{cpu} \times CT\!\left(T_\mathrm{infer}^\mathrm{alg}\right)
\label{eq:alg-cpu-time}    
\end{equation}

where $CT$ denotes the number of CPU core-hours, $T_\mathrm{train}$ is the one-time training time, and $T_\mathrm{infer}$ is the inference time for each problem instance. 

In our experiments, all learning-based approaches used 14 cores during training and inference, resulting in substantial initial core-hour consumption; however, their total core-hour grows slowly when solving problem  instances, owing to nearly constant per-instance inference time. In contrast, metaheuristics (GA and PSO) operated on a single core; however, their cumulative CPU time increased rapidly because their optimisation procedures required a large number of core-hours (even if they could be embarrassingly parallelised over 14 cores). 
The selected crossover examples illustrate this behaviour: \epnco\_SH reaches parity with GA at approximately 1518 problem instances (total core-seconds $\approx$ 146500 s $\approx$ 40.69 h) and with PSO at approximately 1756 problem instances (total core-seconds $\approx$ 149600 s $\approx$ 41.56 h). Similarly, RL\_SH crosses GA at approximately 1800 problem instances (total core-seconds $\approx$ 174800 s $\approx$ 48.56 hours) and PSO at 2087 problem instances (total core-seconds $\approx$ 177900 s $\approx$ 49.42 hours). Models trained on larger instances exhibited proportionally higher crossover points, reflecting longer training times.

Overall, these results demonstrate that despite their higher initial computing investment, learning-based methods become increasingly efficient in core-hours under a fixed compute-budget constraint when repeated large-scale placement decisions are required.

\begin{table}[]
\centering
\caption{Comparison of training and inference time complexities of algorithms.}
\setlength{\tabcolsep}{2pt}
\renewcommand{\arraystretch}{1.05}
{\fontsize{8pt}{9pt}\selectfont
\label{tab:complexity}
\begin{tabular}{lcc}
\hline
\textbf{Algorithms} & \makecell{\textbf{Training Time}\\\textbf{Complexity (Offline)}} & \makecell{\textbf{Inference Time}\\\textbf{Complexity (Online)}} \\
\hline
\epnco~ & $O(E \times T \times M)$ & $O(T \times M)$ \\
RL  & $O(E \times T \times K)$ & $O(T \times K)$ \\
Metaheuristics & None & $O(G \times N \times Cost)$ \\
Heuristics & None & $O(C \times V \times K)$ \\
\hline
\end{tabular}
}
\end{table}

As shown in Table~\ref{tab:complexity}, \epnco~models employ a GNN encoder combined with an RL policy and decoder to learn effective service placement strategies. During training, these models perform multiple RL episodes, resulting in a time complexity of $O(E \times T \times M)$, where $E$ denotes the number of training episodes, $T$ represents the number of placement decisions, and $M$ is the number of edges in the underlying graph. During inference, the trained model generates placement decisions through a single forward pass, resulting in a complexity of $O(T \times M)$.

In contrast, the controlled RL ablation variants models use an MLP-based encoder to evaluate the action space across the available computing nodes at each decision step. Consequently, their training complexity is $O(E \times T \times K)$. Their inference complexity is $O(T \times K)$, where $K$ denotes the number of candidate nodes.

Metaheuristic algorithms do not require an explicit training phase; however, they typically rely on iterative population-based search procedures, resulting in an inference complexity of $O(G \times N \times Cost)$, where $G$ is the number of generations (iterations), $N$ is the population size, and $Cost$ is the computational cost of evaluating one candidate solution. Similarly, rule-based heuristics directly evaluate candidate placements without training, leading to a complexity of $O(C \times V \times K)$, where $C$ represents the number of service components, and $V$ denotes the number of candidate versions for each component.

These results confirm that although learning-based approaches incur a one-time training overhead, the trained models generate placement decisions in approximately one second per instance (0.88--1.03~s for \epnco~and 0.78--0.89~s for the RL ablation variants). The RL variants are slightly faster because their MLP encoder avoids graph message passing. However, this small additional inference cost allows \epnco~to achieve lower aggregate latency costs, and both families remain about two orders of magnitude faster than GA and PSO.

\subsection{Training Analysis}

\begin{figure*}[!t]
    \centering
    \includegraphics[width=\textwidth]{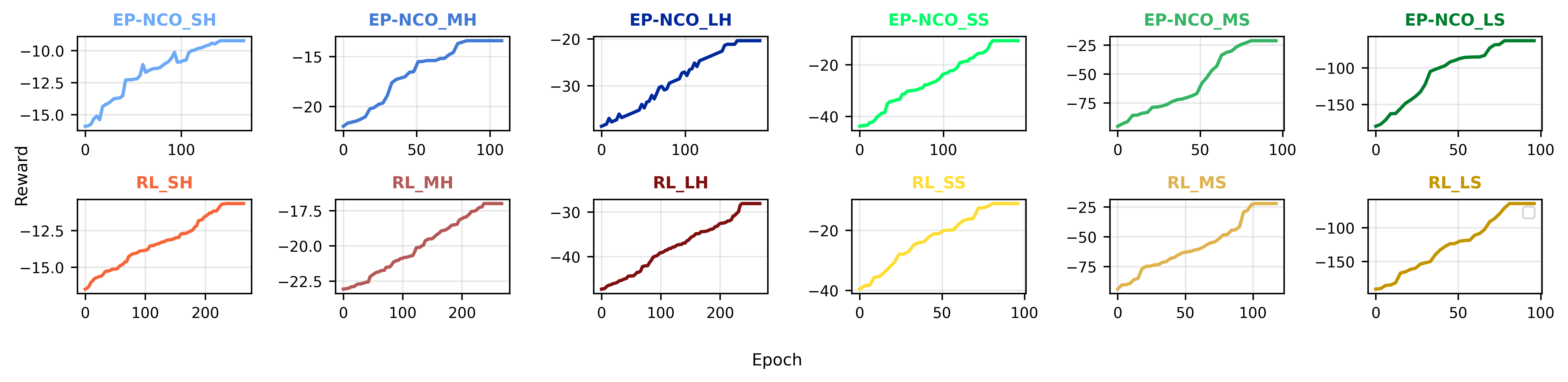}
    \caption{Training reward progression of learning-based placement models. The curves illustrate the convergence behaviour of \epnco~and RL variants across training epochs, highlighting the faster and more stable optimisation achieved by \epnco.}
    \label{fig:reward}
\end{figure*}
Figure \ref{fig:reward} illustrates the reward progression of all learning-based algorithms throughout training, providing insight into their optimisation dynamics and convergence characteristics. A consistent upward trend was observed across the \epnco~family, with rewards steadily improving as training progressed. The hard-decoder variants (\epnco\_SH, \epnco\_MH, and \epnco\_LH) converge more rapidly and exhibit smoother reward trajectories, reflecting the stability introduced by feasibility masking, which prevents the policy from sampling infeasible actions and thus avoids large penalty-driven fluctuations in the reward. In contrast, the soft-decoder versions (\epnco\_SS, \epnco\_MS, and \epnco\_LS) show slightly noisier learning curves but eventually achieve competitive reward levels. This behaviour is expected because the GNN encoder in the \epnco~captures the structural dependencies between services and resources, enabling more globally coherent policy updates during training.

A similar pattern was observed in the controlled RL ablation variants. The RL models employing hard-decoders (RL\_SH, RL\_MH, and RL\_LH) displayed faster and more stable convergence, whereas the soft-decoder variants (RL\_SS, RL\_MS, and RL\_LS) demonstrated higher variance in reward progression, particularly in the early epochs. Because the RL variants rely solely on an MLP encoder without graph-level reasoning, their learning curves tend to exhibit more pronounced fluctuations, particularly for configurations trained on Large instances, where the policy search space becomes substantially wider. Nonetheless, all RL models followed a generally monotonic reward increase, indicating successful training (optimisation), despite their less expressive representations compared to the \epnco-based models.

A comparison of the two families reveals the distinctive role of GNN-based representation learning in accelerating and stabilising optimisation. In particular, \epnco~models consistently achieve higher reward levels within fewer epochs, especially when trained on Small- and Medium-scale instances. This behaviour suggests that they leverage graph structures for more effective credit assignment. RL models, while computationally lighter during training, require more epochs to attain similar reward magnitudes and are more sensitive to the training scale. Overall, the training curves confirm that \epnco~achieves superior convergence behaviour and reward quality, primarily because of its structured GNN encoders (NodeGNN and ServiceGNN) and the more efficient propagation of service--infrastructure dependencies during policy optimisation.

\noindent\textbf{Ablation Insight.}
To further assess the contribution of the proposed architecture, we compared \epnco~with RL ablation variants that replace the GNN encoder with a multilayer perceptron (MLP), a fully connected feedforward network that processes flattened node and service features without exploiting graph structure~\cite{goodfellow2016deep}, and remove explicit dual-graph modelling. The observed performance gap across all evaluations (Figures~\ref{fig:heatmap}--\ref{fig:reward}) indicates that removing structured graph encoding leads to degraded solution quality and less stable training dynamics, highlighting the importance of the relational inductive bias~\cite{battaglia2018relational} in capturing service--infrastructure dependencies.

\subsection{Inference Analysis}

\begin{figure}[]
    \centering
    \includegraphics[width=\linewidth]{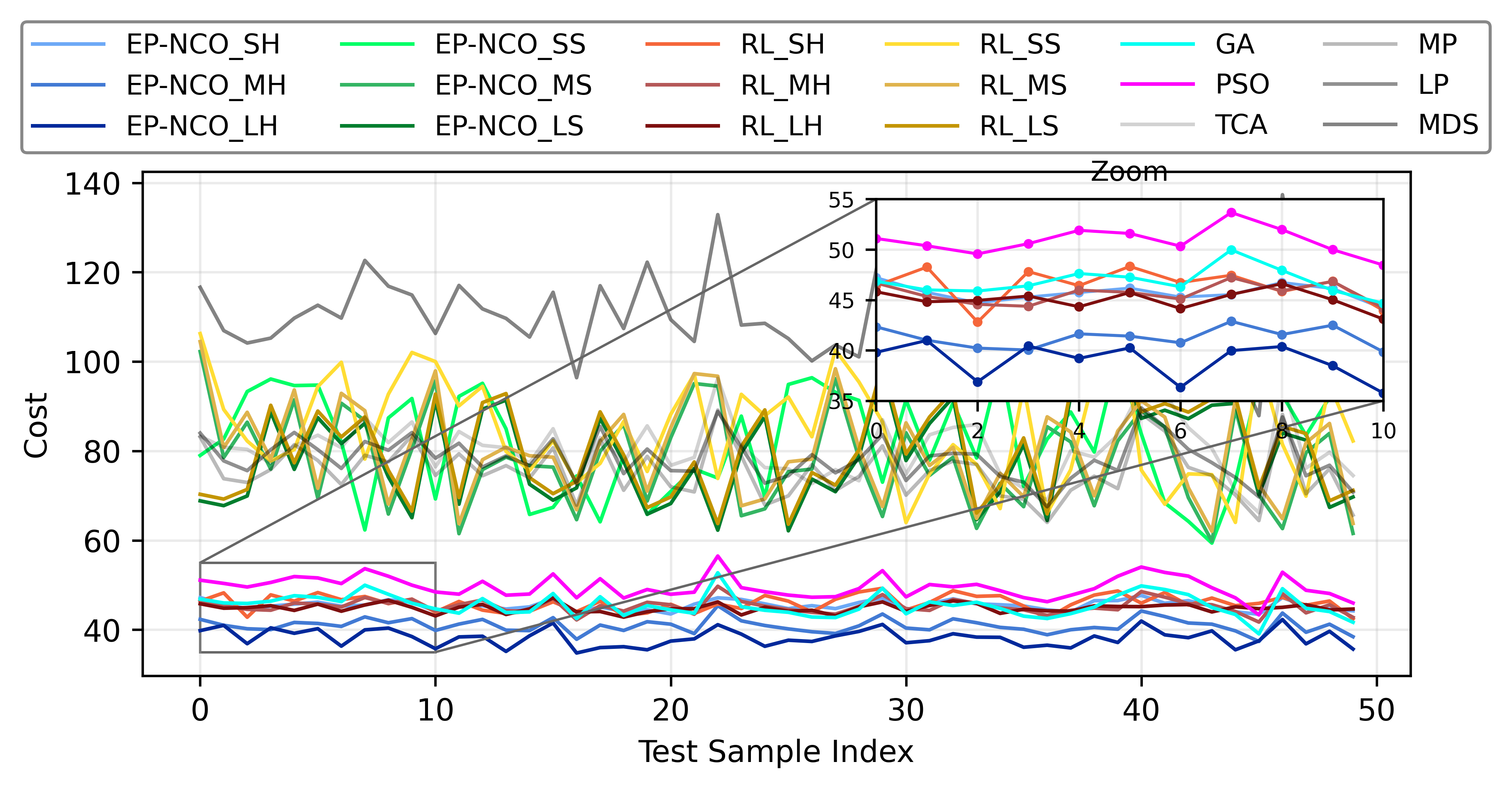}
    \caption{Inference performance on previously unseen XL problem instances. The first 50 test instances are zoomed-in to illustrate the per-instance behaviour of the \epnco~and RL variants compared with the metaheuristic and heuristic baselines. Hard-decoder \epnco~models consistently achieve the lowest and most stable aggregate-latency-cost trajectories.}
    \label{fig:inference}
\end{figure}

To evaluate the generalisation capability of the algorithms, we tested each method on 200 unseen problem instances for each scale. For clearer visual interpretation, Figure \ref{fig:inference} presents the first 50 test instances from the XL configuration, wherein the complexity of the infrastructure renders the performance differences more pronounced. This subset allowed for a detailed comparison of per-execution per instance behaviour while remaining representative of the broader testing distribution, as the overall trends remained consistent across the entire set (200 instances) and across other scales.

Among the learning-based methods, the hard-decoder \epnco~variants consistently achieve the lowest aggregate latency cost with minimal fluctuations between problem instances. Among them, \epnco\_SH and \epnco\_LH exhibited the most stable trajectories, remaining tightly bounded between approximately 35 and 45~s of total per-instance aggregate latency cost. This predictable behaviour stems from the structured GNN encoder within the \epnco, which captures service--infrastructure interactions and yields highly coherent decisions, even for unseen inputs. The soft-decoder variants (\epnco\_SS, \epnco\_MS, and \epnco\_LS) maintained competitive but slightly more variable performance, reflecting the absence of feasibility masking: without it, the policy can select infeasible nodes, which incur penalties and lead to more variable latency costs across instances. The zoomed-in view further highlights this separation: hard-decoder \epnco~models form the lowest-aggregate-latency-cost cluster with minimal oscillations across the 50 tests.

The controlled RL ablation variants demonstrated moderate performance, with consistently higher aggregate latency costs and greater variability than their \epnco~counterparts. RL with hard-decoder variants (RL\_SH, RL\_MH, and RL\_LH) showed smoother curves and improved inference robustness relative to their soft-decoder versions; however, they remained noticeably less stable than \epnco, owing to the absence of a graph-structured inductive bias. RL with soft-decoder models (RL\_SS, RL\_MS, and RL\_LS) exhibited larger fluctuations across problem instances, particularly in the early segments of the curve, indicating less consistent generalisation under high-dimensional input conditions.
Metaheuristics, such as GA and PSO, produced significantly larger and noisier curves, reflecting their lack of learning-based adaptation, where optimisation must be repeated from scratch for each test instance. This resulted in broad variability in solution quality. Rule-based heuristics (TCA, MP, LP, and MDS) remained the least competitive, with MDS consistently producing the highest aggregate latency cost range, often exceeding 100--130 aggregate latency costs across the 50 test instances. These results align with earlier observations: while heuristics are computationally inexpensive, they fail to deliver meaningful optimisation performance for large-scale placements.

Overall, the inference results clearly demonstrate that learning-based methods (particularly \epnco~with hard-decoders) generalise effectively, providing low aggregate latency costs and stable decisions across diverse service placement scenarios. The performance gap between \epnco~and all baseline methods widens further under the XL conditions, highlighting the advantage of combining GNN representations with learned decision policies during inference.

\subsection{Friedman Ranking}
For a fair comparison, all methods were evaluated on the same 200 held-out problem instances per scale, using an identical objective function, cost calculation, and per-service normalisation procedure. Performance was analysed using the non-parametric Friedman test, which is appropriate for multiple-algorithm comparisons without assuming normality and provides a scale-consistent comparative ordering of the evaluated methods \cite{Demar2006StatisticalCO}. For each instance, the algorithms were ranked according to aggregate latency cost (lower is better), and the average rank across all instances yielded the Friedman mean
rank.

The Friedman test indicated statistically significant differences among the algorithms across all scales (all $p<0.001$), rejecting the null hypothesis of equivalent performance and motivating post-hoc comparisons. The Nemenyi post-hoc analysis confirmed that the \epnco~variants significantly outperformed heuristic and RL baselines, whereas the differences with metaheuristics were less consistent across scales. As reported in Table~\ref{tab:friedman-ranking}, the hard-decoder \epnco~variants generally achieved low mean ranks across all problem scales, with \epnco\_LH consistently delivering the strongest performance. In particular, \epnco\_LH achieved both the lowest average aggregate latency cost and the best Friedman mean rank in the Small, Medium, Large, and XLarge settings. \epnco\_MH also performed competitively, obtaining the second-best mean rank at the Small, Medium, and XLarge scales. The hard-decoder RL variants and GA achieved competitive intermediate rankings, whereas the soft-decoder \epnco~and RL variants generally obtained higher mean ranks. In contrast, the rule-based approaches had substantially higher mean ranks, with MDS consistently ranking last. These results highlight the consistent performance advantage and robust scalability of \epnco\_LH as the system complexity increased.

\begin{table}[h]
\centering
\caption{Mean aggregate latency cost (second per service) and Friedman mean rank of all algorithms at each scale.}
\label{tab:friedman-ranking}

\resizebox{1.05\columnwidth}{!}{%
\setlength{\tabcolsep}{2.5pt}
\begin{tabular}{lcccccccc}
\toprule
\multirow{2}{*}{\textbf{Algorithm}}
& \multicolumn{2}{c}{\textbf{Small}}
& \multicolumn{2}{c}{\textbf{Medium}}
& \multicolumn{2}{c}{\textbf{Large}}
& \multicolumn{2}{c}{\textbf{XLarge}} \\

\cmidrule(lr){2-3}
\cmidrule(lr){4-5}
\cmidrule(lr){6-7}
\cmidrule(lr){8-9}

& Mean & Rank
& Mean & Rank
& Mean & Rank
& Mean & Rank \\

\midrule

\epnco\_SH
& 0.618 & 6.03
& 0.609 & 5.85
& 0.623 & 5.95
& 0.604 & 5.53 \\

\epnco\_MH
& 0.513 & 2.07
& 0.547 & 2.18
& 0.575 & 3.28
& 0.543 & 2.04 \\

\epnco\_LH
& \fcolorbox{red}{white}{\textbf{0.511}}
& \fcolorbox{red}{white}{\textbf{1.00}}
& \fcolorbox{red}{white}{\textbf{0.542}}
& \fcolorbox{red}{white}{\textbf{1.01}}
& \fcolorbox{red}{white}{\textbf{0.560}}
& \fcolorbox{red}{white}{\textbf{1.05}}
& \fcolorbox{red}{white}{\textbf{0.540}}
& \fcolorbox{red}{white}{\textbf{1.00}} \\

\epnco\_SS
& 0.750 & 9.17
& 0.759 & 9.00
& 0.780 & 11.72
& 0.753 & 9.00 \\

\epnco\_MS
& 0.782 & 13.33
& 0.803 & 13.02
& 0.789 & 13.25
& 0.813 & 14.00 \\

\epnco\_LS
& 0.753 & 10.18
& 0.761 & 10.00
& 0.780 & 11.64
& 0.755 & 12.00 \\

RL\_SH
& 0.711 & 8.02
& 0.617 & 6.99
& 0.655 & 7.47
& 0.610 & 6.62 \\

RL\_MH
& 0.568 & 4.37
& 0.573 & 4.43
& 0.580 & 4.37
& 0.602 & 4.49 \\

RL\_LH
& 0.541 & 3.18
& 0.560 & 3.27
& 0.567 & 2.20
& 0.596 & 3.40 \\

RL\_SS
& 0.763 & 11.23
& 0.788 & 12.00
& 0.764 & 9.14
& 0.754 & 10.53 \\

RL\_MS
& 0.790 & 14.39
& 0.830 & 14.04
& 0.799 & 14.33
& 0.810 & 13.00 \\

RL\_LS
& 0.765 & 12.24
& 0.780 & 11.00
& 0.765 & 10.14
& 0.754 & 10.47 \\

GA
& 0.584 & 4.65
& 0.585 & 4.63
& 0.596 & 4.43
& 0.605 & 5.13 \\

PSO
& 0.645 & 6.75
& 0.648 & 7.66
& 0.654 & 7.27
& 0.660 & 7.81 \\

TCA
& 0.872 & 15.24
& 1.019 & 16.34
& 0.870 & 15.22
& 1.066 & 16.52 \\

MP
& 0.853 & 14.54
& 0.972 & 15.11
& 0.866 & 14.78
& 1.011 & 15.18 \\

LP
& 0.906 & 16.64
& 1.018 & 16.51
& 0.906 & 16.80
& 1.048 & 16.31 \\

MDS
& 1.183 & 18.00
& 1.398 & 18.00
& 1.191 & 18.00
& 1.466 & 18.00 \\

\bottomrule
\end{tabular}
}
\end{table}

\subsection{Cross-Scale Evaluation Protocol}
Table~\ref{tab:cross-scale-matrix} explicitly summarises the train--test evaluation protocol to further analyse cross-scale generalisation behaviour. The table
highlights whether an \epnco~model trained on a given scale is evaluated on
other scales, making the cross-scale analysis explicit.

\begin{table}[h]
\centering
\caption{Train--test cross-scale evaluation matrix for \epnco.}
\label{tab:cross-scale-matrix}
\setlength{\tabcolsep}{2pt}
\renewcommand{\arraystretch}{1.05}
{\fontsize{8pt}{9pt}\selectfont
\begin{tabular}{|l | c c c c|}
\hline
\diagbox[height=1.7em]{\textbf{Train}}{\textbf{Test}} & Small & Medium & Large & XLarge \\
\hline
Small  & \checkmark & \checkmark & \checkmark & \checkmark \\
Medium & \checkmark & \checkmark & \checkmark & \checkmark \\
Large  & \checkmark & \checkmark & \checkmark & \checkmark \\
\hline
\end{tabular}
}
\end{table}

These comparisons implicitly serve as ablation analysis. 
In particular, the performance gap between the \epnco~and RL baselines highlights the importance of graph-based encoding, whereas the difference between the hard and soft decoders demonstrates the impact of feasibility-aware decision-making.

\section{Conclusion}
\label{sec:conclusion}

This study presents \epnco, an NCO-based solution for latency-aware
microservice placement in heterogeneous edge-cloud infrastructures.
By combining a dual-graph representation with GNN-based encoding and
autoregressive reinforcement learning, \epnco\ models execution,
communication, and bandwidth-sharing effects for reasoning over
resources and dependencies. Experimental evaluations across four
infrastructure scales showed that \epnco\ with hard decoding
outperformed the evaluated baselines. Compared with GA and PSO,
\epnco\_LH reduced mean latency cost by 6.0\%--12.5\%
and 14.4\%--20.8\%, respectively. Compared with the strongest
controlled RL hard-decoder baseline, the reduction was 1.2\%--9.4\%.

On small-scale instances, Gurobi found feasible solutions within
the one-hour time limit but did not prove optimality, achieving a mean
MIP gap of 10.61\%. \epnco\_LH achieved an average
aggregate latency cost of 0.511 s, compared with 0.571 s for the
Gurobi incumbents and an average Gurobi lower bound of 0.509 s. This
result indicates that \epnco\_LH produces high-quality
solutions close, on average, to the solver-derived lower bound.

\epnco\ shows stable performance on the evaluated synthetic test sets.
Its inference time is 0.88--1.03\,s per instance; \epnco\_LH is approximately $99.6\times$ and $87.9\times$ faster
than GA and PSO, respectively. Including training cost,
\epnco\_LH reaches wall-clock break-even after
approximately 994 and 1,128 instances relative to GA and PSO, while
\epnco\_SH reaches break-even after approximately 95 and
108 instances. The model assumes a static topology and direct-link
communication and relies on synthetic workloads, which may limit its
applicability in dynamic real-world environments. Nevertheless, these
findings suggest that \epnco\ provides a scalable and efficient
solution for latency-sensitive edge-cloud applications.

Future work will investigate dynamic workloads, multi-objective
optimisation of latency, reliability, and energy consumption, and
federated training for decentralised edge systems.

\begin{acknowledgements}
This research is supported in part by the Engineering and Physical Sciences Research Council (EPSRC), UK Research and Innovation (UKRI), under Grant EP/Y028813/1. Additional support is provided by the Knowledge Foundation of Sweden (KKS).
\end{acknowledgements}

\printcredits
\section*{Data availability}
The synthetic problem instances (Small, Medium, Large and XLarge training and test sets) used in this study are publicly available at \url{https://drive.google.com/drive/folders/1EBjEOiklv3lFB4W6p8aBOZsQx5tJC7bx}. The source code of \epnco, is available at \cite{epnco_code}.

\section*{Declaration of competing interest}
The authors declare that they have no known competing financial interests or personal relationships that could have appeared to influence the work reported in this paper.

\section*{Declaration of generative AI and AI-assisted technologies in the manuscript preparation process}
During the preparation of this work, the authors used ChatGPT (OpenAI) to improve the language and readability of the manuscript. After using this tool, the authors reviewed and edited the content as needed and take full responsibility for the content of the published article.
\bibliographystyle{unsrtnat}
\bibliography{ref}

@article{taleb2025survey,
  author  = {Taleb, Imane and Guillaume, Jean-Loup and Duthil, Benjamin},
  title   = {A Survey on Services Placement Algorithms in Integrated Cloud-Fog/Edge Computing},
  journal = {ACM Computing Surveys},
  volume  = {57},
  number  = {11},
  pages   = {1--36},
  year    = {2025}
}

@article{supraja2025ai,
  author  = {Supraja, Thatikonda and Chawla, Priyanka and Gill, Sukhpal Singh},
  title   = {{AI}-Driven Service Placement in Fog and Edge Computing Environments: A Systematic Review, Taxonomy and Future Directions},
  journal = {Cluster Computing},
  volume  = {28},
  number  = {16},
  pages   = {1--39},
  year    = {2025},
  doi     = {10.1007/s10586-025-05759-2}
}

@article{kong2022edge,
  author  = {Kong, Linghe and Tan, Jinlin and Huang, Junqin and Chen, Guihai and Wang, Shuaitian and Jin, Xi and Zeng, Peng and Khan, Muhammad and Das, Sajal K.},
  title   = {Edge-Computing-Driven {Internet of Things}: A Survey},
  journal = {ACM Computing Surveys},
  volume  = {55},
  number  = {8},
  pages   = {1--41},
  year    = {2022}
}

@article{10.1145/3391196,
  author  = {Salaht, Farah A{\"\i}t and Desprez, Fr{\'e}d{\'e}ric and Lebre, Adrien},
  title   = {An Overview of Service Placement Problem in Fog and Edge Computing},
  journal = {ACM Computing Surveys},
  volume  = {53},
  number  = {3},
  pages   = {1--35},
  year    = {2020},
  doi     = {10.1145/3391196}
}

@article{malazi2022dynamic,
  author  = {Tabatabaee Malazi, Hadi and Chaudhry, Saqib Rasool and Kazmi, Aqeel and Palade, Andrei and Cabrera, Christian and White, Gary and Clarke, Siobh{\'a}n},
  title   = {Dynamic Service Placement in Multi-Access Edge Computing: A Systematic Literature Review},
  journal = {IEEE Access},
  volume  = {10},
  pages   = {32639--32688},
  year    = {2022},
  doi     = {10.1109/ACCESS.2022.3160738}
}

@article{pallewatta2023taxonomy,
  author  = {Pallewatta, Samodha and Kostakos, Vassilis and Buyya, Rajkumar},
  title   = {Placement of Microservices-Based {IoT} Applications in Fog Computing: A Taxonomy and Future Directions},
  journal = {ACM Computing Surveys},
  volume  = {55},
  number  = {14s},
  pages   = {1--43},
  year    = {2023},
  doi     = {10.1145/3592598}
}

@article{sonkoly2021survey,
  author  = {Sonkoly, Bal{\'a}zs and Czentye, J{\'a}nos and Szalay, M{\'a}rk and N{\'e}meth, Bal{\'a}zs and Toka, L{\'a}szl{\'o}},
  title   = {Survey on Placement Methods in the Edge and Beyond},
  journal = {IEEE Communications Surveys \& Tutorials},
  volume  = {23},
  number  = {4},
  pages   = {2590--2629},
  year    = {2021}
}

@inproceedings{taneja2017resource,
  author    = {Taneja, Mohit and Davy, Alan},
  title     = {Resource Aware Placement of {IoT} Application Modules in Fog-Cloud Computing Paradigm},
  booktitle = {Proc. IFIP/IEEE Symposium on Integrated Network and Service Management (IM)},
  pages     = {1222--1228},
  year      = {2017},
  doi       = {10.23919/INM.2017.7987464}
}

@article{mahmud2018latency,
  author  = {Mahmud, Redowan and Ramamohanarao, Kotagiri and Buyya, Rajkumar},
  title   = {Latency-Aware Application Module Management for Fog Computing Environments},
  journal = {ACM Transactions on Internet Technology},
  volume  = {19},
  number  = {1},
  pages   = {1--21},
  year    = {2018},
  doi     = {10.1145/3186592}
}

@article{gupta2017ifogsim,
  author  = {Gupta, Harshit and Vahid Dastjerdi, Amir and Ghosh, Soumya K. and Buyya, Rajkumar},
  title   = {{iFogSim}: A Toolkit for Modeling and Simulation of Resource Management Techniques in the {Internet of Things}, Edge and Fog Computing Environments},
  journal = {Software: Practice and Experience},
  volume  = {47},
  number  = {9},
  pages   = {1275--1296},
  year    = {2017},
  doi     = {10.1002/spe.2509}
}

@article{brogi2017qos,
  author  = {Brogi, Antonio and Forti, Stefano},
  title   = {{QoS}-Aware Deployment of {IoT} Applications Through the Fog},
  journal = {IEEE Internet of Things Journal},
  volume  = {4},
  number  = {5},
  pages   = {1185--1192},
  year    = {2017},
  doi     = {10.1109/JIOT.2017.2701408}
}

@article{pallewatta2022qos,
  author  = {Pallewatta, Samodha and Kostakos, Vassilis and Buyya, Rajkumar},
  title   = {{QoS}-Aware Placement of Microservices-Based {IoT} Applications in Fog Computing Environments},
  journal = {Future Generation Computer Systems},
  volume  = {131},
  pages   = {121--136},
  year    = {2022},
  doi     = {10.1016/j.future.2022.01.012}
}

@article{pallewatta2024reliability,
  author  = {Pallewatta, Samodha and Kostakos, Vassilis and Buyya, Rajkumar},
  title   = {Reliability-Aware Proactive Placement of Microservices-Based {IoT} Applications in Fog Computing Environments},
  journal = {IEEE Transactions on Mobile Computing},
  volume  = {23},
  number  = {12},
  pages   = {11326--11342},
  year    = {2024},
  doi     = {10.1109/TMC.2024.3394486}
}

@inproceedings{mahjoubi2021optimal,
  author    = {Mahjoubi, Ayeh and Taheri, Javid and Grinnemo, Karl-Johan and Deng, Shuiguang},
  title     = {Optimal Placement of Recurrent Service Chains on Distributed Edge-Cloud Infrastructures},
  booktitle = {Proc. IEEE 46th Conference on Local Computer Networks (LCN)},
  pages     = {495--502},
  year      = {2021}
}

@inproceedings{herabad2024optimizing,
  author    = {Garshasbi Herabad, Mohammadsadeq and Taheri, Javid and Ahmed, Bestoun S. and Curescu, Calin},
  title     = {Optimizing Service Placement in Edge-to-Cloud {AR/VR} Systems Using a Multi-Objective Genetic Algorithm},
  booktitle = {Proc. 14th International Conference on Cloud Computing and Services Science (CLOSER)},
  year      = {2024}
}

@article{herabad2025psoga,
  author  = {Garshasbi Herabad, Mohammadsadeq and Taheri, Javid and Ahmed, Bestoun S. and Curescu, Calin},
  title   = {{E-PSOGA}: An Enhanced Hybrid Metaheuristic for Optimal Edge-to-Cloud Placement of Services With Multi-Version Components},
  journal = {IEEE Access},
  year    = {2025},
  doi     = {10.1109/ACCESS.2025.3603329}
}

@article{electronics15010065,
  author  = {Garshasbi Herabad, Mohammadsadeq and Taheri, Javid and Ahmed, Bestoun S. and Curescu, Calin},
  title   = {A Lightweight Learning-Based Approach for Online Edge-to-Cloud Service Placement},
  journal = {Electronics},
  volume  = {15},
  number  = {1},
  pages   = {65},
  year    = {2026},
  doi     = {10.3390/electronics15010065}
}

@article{apat2024hybrid,
  author  = {Apat, Hemant Kumar and Sahoo, Bibhudutta and Goswami, Veena and Barik, Rabindra K.},
  title   = {A Hybrid Meta-Heuristic Algorithm for Multi-Objective {IoT} Service Placement in Fog Computing Environments},
  journal = {Decision Analytics Journal},
  volume  = {10},
  pages   = {100379},
  year    = {2024}
}

@article{bey2024quantum,
  author  = {Bey, Marlom and Kuila, Pratyay and Naik, Banavath Balaji and Ghosh, Santanu},
  title   = {Quantum-Inspired Particle Swarm Optimization for Efficient {IoT} Service Placement in Edge Computing Systems},
  journal = {Expert Systems with Applications},
  volume  = {236},
  pages   = {121270},
  year    = {2024}
}

@article{huang2020ant,
  author  = {Huang, Tiansheng and Lin, Weiwei and Xiong, Chennian and Pan, Rui and Huang, Jingxuan},
  title   = {An Ant Colony Optimization-Based Multiobjective Service Replicas Placement Strategy for Fog Computing},
  journal = {IEEE Transactions on Cybernetics},
  volume  = {51},
  number  = {11},
  pages   = {5595--5608},
  year    = {2021}
}

@article{khan2022code,
  author  = {Khan, Muhammad Asif and Baccour, Emna and Erbad, Aiman and Hamila, Ridha and Hamdi, Mounir},
  title   = {{CODE}: Computation Offloading in {D2D}-Edge System for Video Streaming},
  journal = {IEEE Systems Journal},
  volume  = {17},
  number  = {3},
  pages   = {4014--4025},
  year    = {2023}
}

@article{wu2021online,
  author  = {Wu, Chunrong and Peng, Qinglan and Xia, Yunni and Ma, Yong and Zheng, Wangbo and Xie, Hong and Pang, Shanchen and Li, Fan and Fu, Xiaodong and Li, Xiaobo and others},
  title   = {Online User Allocation in Mobile Edge Computing Environments: A Decentralized Reactive Approach},
  journal = {Journal of Systems Architecture},
  volume  = {113},
  pages   = {101904},
  year    = {2021}
}

@article{fahimullah2024machine,
  author  = {Fahimullah, Muhammad and Ahvar, Shohreh and Agarwal, Mihir and Trocan, Maria},
  title   = {Machine Learning-Based Solutions for Resource Management in Fog Computing},
  journal = {Multimedia Tools and Applications},
  volume  = {83},
  number  = {8},
  pages   = {23019--23045},
  year    = {2024}
}

@article{lingayya2024dynamic,
  author  = {Lingayya, Sadananda and Jodumutt, Sathyendra Bhat and Pawar, Sanjay Rangrao and Vylala, Anoop and Chandrasekaran, Senthilkumar},
  title   = {Dynamic Task Offloading for Resource Allocation and Privacy-Preserving Framework in {KubeEdge}-Based Edge Computing Using Machine Learning},
  journal = {Cluster Computing},
  volume  = {27},
  number  = {7},
  pages   = {9415--9431},
  year    = {2024}
}

@article{clifton2020q,
  author  = {Clifton, Jesse and Laber, Eric},
  title   = {{Q}-Learning: Theory and Applications},
  journal = {Annual Review of Statistics and Its Application},
  volume  = {7},
  pages   = {279--301},
  year    = {2020},
  doi     = {10.1146/annurev-statistics-031219-041220}
}

@article{jang2019q,
  author  = {Jang, Byungjin and Kim, Minho and Harerimana, Gaspard and Kim, Jae Wook},
  title   = {{Q}-Learning Algorithms: A Comprehensive Classification and Applications},
  journal = {IEEE Access},
  volume  = {7},
  pages   = {133653--133667},
  year    = {2019},
  doi     = {10.1109/ACCESS.2019.2941229}
}

@article{liu2022deep,
  author  = {Liu, Tao and Ni, Shuai and Li, Xiang and Zhu, Yan and Kong, Linghe},
  title   = {Deep Reinforcement Learning Based Approach for Online Service Placement and Computation Resource Allocation in Edge Computing},
  journal = {IEEE Transactions on Mobile Computing},
  year    = {2022},
  doi     = {10.1109/TMC.2022.3141230}
}

@article{chen2022joint,
  author  = {Chen, Yao and Sun, Yang and Yang, Bo and Taleb, Tarik},
  title   = {Joint Caching and Computing Service Placement for Edge-Enabled {IoT} Based on Deep Reinforcement Learning},
  journal = {IEEE Internet of Things Journal},
  volume  = {9},
  number  = {20},
  pages   = {20006--20017},
  year    = {2022},
  doi     = {10.1109/JIOT.2022.3159913}
}

@article{afachao2024efficient,
  author  = {Afachao, Kevin and Abu-Mahfouz, Adnan M. and Hanke, Gerhard P.},
  title   = {Efficient Microservice Deployment in the Edge-Cloud Networks With Policy-Gradient Reinforcement Learning},
  journal = {IEEE Access},
  volume  = {12},
  pages   = {133110--133124},
  year    = {2024},
  doi     = {10.1109/ACCESS.2024.3461149}
}

@article{santos2025hephaestus,
  author  = {Santos, Jos{\'e} and Zaccarini, Mattia and Poltronieri, Filippo and Tortonesi, Mauro and Stefanelli, Cesare and Di Cicco, Nicola and De Turck, Filip},
  title   = {{HephaestusForge}: Optimal Microservice Deployment Across the Compute Continuum via Reinforcement Learning},
  journal = {Future Generation Computer Systems},
  volume  = {166},
  pages   = {107680},
  year    = {2025},
  doi     = {10.1016/j.future.2024.107680}
}

@article{khansari2024scalable,
  author  = {Khansari, Mina Emami and Sharifian, Saeed},
  title   = {A Scalable Modified Deep Reinforcement Learning Algorithm for Serverless {IoT} Microservice Composition Infrastructure in Fog Layer},
  journal = {Future Generation Computer Systems},
  volume  = {153},
  pages   = {206--221},
  year    = {2024},
  doi     = {10.1016/j.future.2023.11.022}
}

@article{lv2023graph,
  author  = {Lv, Wenkai and Yang, Pengfei and Zheng, Tianyang and Lin, Chengmin and Wang, Zhenyi and Deng, Minwen and Wang, Quan},
  title   = {Graph-Reinforcement-Learning-Based Dependency-Aware Microservice Deployment in Edge Computing},
  journal = {IEEE Internet of Things Journal},
  volume  = {11},
  number  = {1},
  pages   = {1604--1615},
  year    = {2024},
  doi     = {10.1109/JIOT.2023.3289228}
}

@article{chen2024gnnaided,
  author  = {Chen, Shuangwu and Yuan, Qifeng and Li, Jiangming and He, Huasen and Li, Sen and Jiang, Xiaofeng and Yang, Jian},
  title   = {Graph Neural Network Aided Deep Reinforcement Learning for Microservice Deployment in Cooperative Edge Computing},
  journal = {IEEE Transactions on Services Computing},
  volume  = {17},
  number  = {6},
  pages   = {3742--3757},
  year    = {2024},
  doi     = {10.1109/TSC.2024.3417241}
}

@article{chen2024adaptive,
  author  = {Chen, Lixing and Bai, Yang and Zhou, Pan and Li, Youqi and Qu, Zhe and Xu, Jie},
  title   = {On Adaptive Edge Microservice Placement: A Reinforcement Learning Approach Endowed With Graph Comprehension},
  journal = {IEEE Transactions on Mobile Computing},
  volume  = {23},
  number  = {12},
  pages   = {11144--11158},
  year    = {2024},
  doi     = {10.1109/TMC.2024.3396510}
}

@article{sharma2024intelligent,
  author  = {Sharma, Ankur and Thangaraj, Veni},
  title   = {Intelligent Service Placement Algorithm Based on {DDQN} and Prioritized Experience Replay in {IoT}-Fog Computing Environment},
  journal = {Internet of Things},
  volume  = {25},
  pages   = {101112},
  year    = {2024}
}

@misc{garg2021heuristicreinforcementlearningalgorithms,
  author        = {Garg, Dhruv and Narendra, Nanjangud C. and Tesfatsion, Selome},
  title         = {Heuristic and Reinforcement Learning Algorithms for Dynamic Service Placement on Mobile Edge Cloud},
  year          = {2021},
  eprint        = {2111.00240},
  archivePrefix = {arXiv},
  primaryClass  = {cs.NI},
  howpublished  = {arXiv preprint arXiv:2111.00240},
  url           = {https://arxiv.org/abs/2111.00240}
}

@article{bengio2021methodological,
  author  = {Bengio, Yoshua and Lodi, Andrea and Prouvost, Antoine},
  title   = {Machine Learning for Combinatorial Optimization: A Methodological Tour d'Horizon},
  journal = {European Journal of Operational Research},
  volume  = {290},
  number  = {2},
  pages   = {405--421},
  year    = {2021},
  doi     = {10.1016/j.ejor.2018.10.063}
}

@article{mazyavkina2021rlco,
  author  = {Mazyavkina, Nina and Sviridov, Sergey and Ivanov, Sergei and Burnaev, Evgeny},
  title   = {Reinforcement Learning for Combinatorial Optimization: A Survey},
  journal = {Computers \& Operations Research},
  volume  = {134},
  pages   = {105400},
  year    = {2021},
  doi     = {10.1016/j.cor.2021.105400}
}

@article{vesselinova2020learning,
  author  = {Vesselinova, Natalia and Steinert, Rebecca and P{\'e}rez-Ram{\'i}rez, Daniel F. and Boman, Magnus},
  title   = {Learning Combinatorial Optimization on Graphs: A Survey With Applications to Networking},
  journal = {IEEE Access},
  volume  = {8},
  pages   = {120388--120416},
  year    = {2020},
  doi     = {10.1109/ACCESS.2020.3005682}
}

@article{chung2025neural,
  author  = {Chung, Ka Tai and Lee, C. K. M. and Tsang, Y. P.},
  title   = {Neural Combinatorial Optimization With Reinforcement Learning in Industrial Engineering: A Survey},
  journal = {Artificial Intelligence Review},
  year    = {2025},
  doi     = {10.1007/s10462-024-11045-1}
}

@article{jiang2020neural,
  author  = {Jiang, Qingmiao and Zhang, Yuan and Yan, Jinyao},
  title   = {Neural Combinatorial Optimization for Energy-Efficient Offloading in Mobile Edge Computing},
  journal = {IEEE Access},
  volume  = {8},
  pages   = {35077--35089},
  year    = {2020}
}

@article{xiao2025nco,
  author  = {Xiao, Xiang-Jie and Wang, Yong and Huang, Pei-Qiu and Wang, Kezhi},
  title   = {Neural Combinatorial Optimization for Multiobjective Task Offloading in Mobile Edge Computing},
  journal = {IEEE Transactions on Vehicular Technology},
  volume  = {74},
  number  = {7},
  pages   = {10869--10880},
  year    = {2025},
  doi     = {10.1109/TVT.2025.3546914}
}

@book{goodfellow2016deep,
  author    = {Goodfellow, Ian and Bengio, Yoshua and Courville, Aaron},
  title     = {Deep Learning},
  publisher = {MIT Press},
  address   = {Cambridge, MA, USA},
  year      = {2016}
}

@misc{battaglia2018relational,
  author       = {Battaglia, Peter W. and Hamrick, Jessica B. and Bapst, Victor and others},
  title        = {Relational Inductive Biases, Deep Learning, and Graph Networks},
  howpublished = {arXiv preprint arXiv:1806.01261},
  year         = {2018}
}

@article{Demar2006StatisticalCO,
  author  = {Dem{\v{s}}ar, Janez},
  title   = {Statistical Comparisons of Classifiers over Multiple Data Sets},
  journal = {Journal of Machine Learning Research},
  volume  = {7},
  pages   = {1--30},
  year    = {2006}
}

@misc{epnco_code,
  author       = {Abedpour, Kimia},
  title        = {{EP-NCO}: Latency-Aware Service Placement Using Neural Combinatorial Optimisation (Code)},
  howpublished = {\url{https://github.com/kimiaa45-ab/EP-NCO}},
  year         = {2026},
  note         = {Accessed May 2026}
}


\bio{img/kimia.png}
\textbf{Kimia Abedpour} received her B.Sc. and M.Sc. degrees in Computer Engineering (Software) from Marlik Nowshahr Institute and Tabarestan Chalus Institute, Iran, in 2019 and 2021, respectively. She is currently pursuing her Ph.D. in Computer Science at Queen's University Belfast, United Kingdom. Her research interests include cloud and edge/fog computing, service placement, and AI-based optimisation techniques.
\endbio
\vspace{1em}

\bio{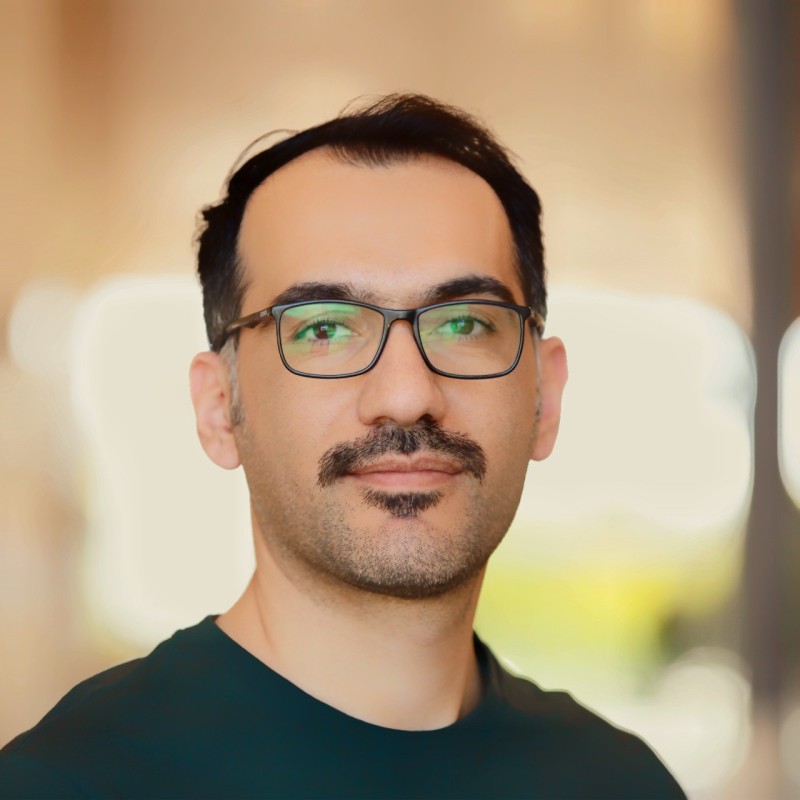}
\textbf{Mohammadsadeq Garshasbi Herabad} is a PhD researcher in Computer Science at Karlstad University, Sweden. His research focuses on intelligent and adaptive systems for distributed and dynamic computing environments. His work investigates the use of machine learning and optimisation techniques to improve system performance and reliability under dynamic conditions. He has authored multiple peer-reviewed publications in international venues. His research interests include distributed systems, edge computing, and AI-driven optimisation.
\endbio
\vspace{1em}

\bio{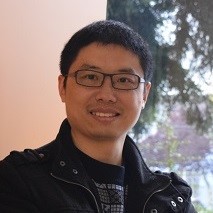}
\textbf{Zheng Li} received his Ph.D. and M.Phil. degrees from the Australian National University (ANU) and the University of New South Wales (UNSW), respectively. He was a graduate researcher with the Software Systems Research Group at National ICT Australia (NICTA). He is currently a Lecturer at the School of Electronics, Electrical Engineering, and Computer Science, Queen's University Belfast, UK. His research interests include big data analytics, edge/cloud computing, empirical software engineering, and performance engineering.
\endbio
\vspace{1em}

\bio{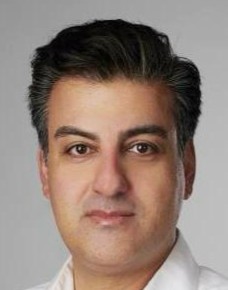}
\textbf{Javid Taheri} is a full Professor with the School of Electronics, Electrical Engineering, and Computer Science, Queen's University Belfast (UK), and with the Department of Mathematics and Computer Science, Karlstad University (Sweden). He was also a visiting professor at Ericsson HQ (Sweden) in 2022--2025. He received his Ph.D. in Mobile Computing from the University of Sydney, Australia. His research interests include cloud and edge computing, network virtualisation, software-defined networking, and AI-based optimisation.
\endbio
\end{document}